\input harvmac.tex
\noblackbox
\input epsf.sty

\font\cmss=cmss10 \font\cmsss=cmss10 at 7pt

\def\IB{\relax\hbox{$\inbar\kern-.3em{\rm B}$}}
\def\IC{\relax\hbox{$\inbar\kern-.3em{\rm C}$}}
\def\IQ{\relax\hbox{$\inbar\kern-.3em{\rm Q}$}}
\def\ID{\relax\hbox{$\inbar\kern-.3em{\rm D}$}}
\def\IE{\relax\hbox{$\inbar\kern-.3em{\rm E}$}}
\def\IF{\relax\hbox{$\inbar\kern-.3em{\rm F}$}}
\def\IG{\relax\hbox{$\inbar\kern-.3em{\rm G}$}}
\def\IGa{\relax\hbox{${\rm I}\kern-.18em\Gamma$}}
\def\IH{\relax{\rm I\kern-.18em H}}
\def\IK{\relax{\rm I\kern-.18em K}}
\def\IL{\relax{\rm I\kern-.18em L}}
\def\IP{\relax{\rm I\kern-.18em P}}
\def\IR{\relax{\rm I\kern-.18em R}}
\def\Z{\relax\ifmmode\mathchoice{\hbox{\cmss Z\kern-.4em Z}}{\hbox{\cmss Z\kern-.4em Z}} {\lower.9pt\hbox{\cmsss Z\kern-.4em Z}}{\lower1.2pt\hbox{\cmsss Z\kern-.4em Z}}\else{\cmss Z\kern-.4em Z}\fi}

\def\II{\relax{\rm I\kern-.18em I}}
\def\one{\relax{\rm 1\kern-.25em I}}

\def\CLL{\relax{\CL\kern-.74em \CL}}

\def\CA {{\cal A}}
\def\CB {{\cal B}}

\def\CF {{\cal F}}

\def\CL {{\cal L}}

\def\CN {{\cal N}}

\def\CW {{\cal W}}


\def\tilde{\widetilde}

\def\bar{\overline}


\def\bar{\overline}

\def\det{{\rm det}}

\def\Tr{{\rm Tr}}

\def\IH{{\bf H}}

\def\vol{{\rm vol}}

\def\half{{1\over 2}}


\lref\toappear{M.~Aganagic, C.~Vafa, Work in progress.}. 

\lref\strominger{
  J.~Polchinski and A.~Strominger,
  ``New Vacua for Type II String Theory,''
  Phys.\ Lett.\  B {\bf 388}, 736 (1996)
  [arXiv:hep-th/9510227].
}

\lref\KS{
  I.~R.~Klebanov and M.~J.~Strassler,
  ``Supergravity and a confining gauge theory: Duality cascades and
  chiSB-resolution of naked singularities,''
  JHEP {\bf 0008}, 052 (2000)
  [arXiv:hep-th/0007191].
}

\lref\seiberg{N. Seiberg, ``Electric-Magnetic Duality in 
  Supersymmetric Non-Abelian Gauge Theories'',
  Nucl.Phys. B435 (1995) 129-146, arXiv:hep-th/9411149v1
} 

\lref\KachruVJ{
  B.~Florea, S.~Kachru, J.~McGreevy and N.~Saulina,
  ``Stringy instantons and quiver gauge theories,''
  JHEP {\bf 0705}, 024 (2007)
  [arXiv:hep-th/0610003].
}

\lref\Diaconescu{
  D.~E.~Diaconescu, R.~Donagi and B.~Florea,
  ``Metastable quivers in string compactifications,''
  Nucl.\ Phys.\  B {\bf 774}, 102 (2007)
  [arXiv:hep-th/0701104].
}

\lref\Aganagic{
  M.~Aganagic, C.~Beem, J.~Seo and C.~Vafa,
  ``Geometrically induced metastability and holography,''
  [arXiv:hep-th/0610249].
}

\lref\Seo{
  J.~J.~Heckman, J.~Seo and C.~Vafa,
  ``Phase structure of a brane/anti-brane system at large $N$,''
  [arXiv:hep-th/0702077].
}

\lref\Heckman{
  J.~J.~Heckman and C.~Vafa,
  ``Geometrically Induced Phase Transitions at Large N,''
  arXiv:0707.4011 [hep-th].
}

\lref\KachruVJ{
  S.~Kachru and J.~McGreevy,
  ``Supersymmetric three-cycles and (super)symmetry breaking,''
  Phys.\ Rev.\  D {\bf 61}, 026001 (2000)
  [arXiv:hep-th/9908135].
}

\lref\VafaWI{
  C.~Vafa,
  ``Superstrings and topological strings at large N,''
  J.\ Math.\ Phys.\  {\bf 42}, 2798 (2001)
  [arXiv:hep-th/0008142].
}

\lref\KachruGS{
  S.~Kachru, J.~Pearson and H.~L.~Verlinde,
  ``Brane/flux annihilation and the string dual of a non-supersymmetric field
  theory,''
  JHEP {\bf 0206}, 021 (2002)
  [arXiv:hep-th/0112197].
}

\lref\KachruAW{
  S.~Kachru, R.~Kallosh, A.~Linde and S.~P.~Trivedi,
  ``De Sitter vacua in string theory,''
  Phys.\ Rev.\  D {\bf 68}, 046005 (2003)
  [arXiv:hep-th/0301240].
}

\lref\ArgurioNY{
  R.~Argurio, M.~Bertolini, S.~Franco and S.~Kachru,
  ``Gauge/gravity duality and meta-stable dynamical supersymmetry breaking,''
  JHEP {\bf 0701}, 083 (2007)
  [arXiv:hep-th/0610212].
}

\lref\VerlindeBC{
  H.~Verlinde,
  ``On metastable branes and a new type of magnetic monopole,''
  [arXiv:hep-th/0611069].
}

\lref\ArgurioQK{
  R.~Argurio, M.~Bertolini, S.~Franco and S.~Kachru,
  ``Metastable vacua and D-branes at the conifold,''
  JHEP {\bf 0706}, 017 (2007)
  [arXiv:hep-th/0703236].
}

\lref\MurthyQM{
  S.~Murthy,
  ``On supersymmetry breaking in string theory from gauge theory in a throat,''
  [arXiv:hep-th/0703237].
}

\lref\MarsanoFE{
  J.~Marsano, K.~Papadodimas and M.~Shigemori,
  ``Non-supersymmetric brane / anti-brane configurations in type IIA and M theory,''
  arXiv:0705.0983 [hep-th].
}

\lref\douglas{
  M.~R.~Douglas, J.~Shelton and G.~Torroba,
  ``Warping and supersymmetry breaking,''
  arXiv:0704.4001 [hep-th].
}

\lref\ISS{
  K.~Intriligator, N.~Seiberg and D.~Shih,
  ``Dynamical SUSY breaking in meta-stable vacua,''
  JHEP {\bf 0604}, 021 (2006)
  [arXiv:hep-th/0602239].
}

\lref\OoguriPJ{
  H.~Ooguri and Y.~Ookouchi,
  ``Landscape of supersymmetry breaking vacua in geometrically realized gauge theories,''
  Nucl.\ Phys.\  B {\bf 755}, 239 (2006)
  [arXiv:hep-th/0606061].
}

\lref\OoguriBG{
  H.~Ooguri and Y.~Ookouchi,
  ``Meta-stable supersymmetry breaking vacua on intersecting branes,''
  Phys.\ Lett.\  B {\bf 641}, 323 (2006)
  [arXiv:hep-th/0607183].
}

\lref\FrancoHT{
  S.~Franco, I.~Garcia-Etxebarria and A.~M.~Uranga,
  ``Non-supersymmetric meta-stable vacua from brane configurations,''
  JHEP {\bf 0701}, 085 (2007)
  [arXiv:hep-th/0607218].
}

\lref\Bena{
  I.~Bena, E.~Gorbatov, S.~Hellerman, N.~Seiberg and D.~Shih,
  ``A note on (meta)stable brane configurations in MQCD,''
  JHEP {\bf 0611}, 088 (2006)
  [arXiv:hep-th/0608157].
}

\lref\Ahn{
C.~Ahn, ``Brane configurations for nonsupersymmetric meta-stable
  vacua in SQCD with adjoint matter,'' Class.\ Quant.\ Grav.\ {\bf
  24}, 1359 (2007) [arXiv:hep-th/0608160];
C.~Ahn,``M-theory lift of meta-stable brane configuration in
  symplectic and
  orthogonal gauge groups,''
  Phys.\ Lett.\  B {\bf 647}, 493 (2007),
  [arXiv:hep-th/0610025];
C.~Ahn,
  ``Meta-stable brane configuration with orientifold 6 plane,''
  JHEP {\bf 0705}, 053 (2007), 
  [arXiv:hep-th/0701145];
C.~Ahn,
  ``More on Meta-Stable Brane Configuration,'' 
  Class.\ Quant.\ Grav.\ {\bf 24}, 3603 (2007),
  [arXiv:hep-th/0702038];
C.~Ahn, 
  ``Meta-Stable Brane Configuration and Gauged Flavor Symmetry,'' 
  [arXiv:hep-th/0703015];
C.~Ahn,
  ``Meta-Stable Brane Configuration of Product Gauge Groups,'' 
  arXiv:0704.0121 [hep-th];
C.~Ahn, 
  ``Meta-Stable Brane Configurations with Five NS5-Branes,''
  arXiv:0705.0056 [hep-th];
C.~Ahn,
  ``Meta-Stable Brane Configurations by Adding an 
  Orientifold-Plane to Giveon-Kutasov,'' 
  arXiv:0706.0042 [hep-th];
C.~Ahn,
  ``More Meta-Stable Brane Configurations without D6-Branes,''
  arXiv:0707.0092 [hep-th].
}

\lref\TatarDM{
  R.~Tatar and B.~Wetenhall,
  ``Metastable vacua, geometrical engineering and MQCD transitions,''
  JHEP {\bf 0702}, 020 (2007)
  [arXiv:hep-th/0611303].
}

\lref\GiveonFK{
  A.~Giveon and D.~Kutasov,
  ``Gauge symmetry and supersymmetry breaking 
  from intersecting branes,''
  [arXiv:hep-th/0703135].
}

\lref\TatarZA{
  R.~Tatar and B.~Wetenhall,
  ``Metastable Vacua and Complex Deformations,''
  \vskip.01pt  
  arXiv:0707.2712 [hep-th].
}

\lref\HoravaVS{
  P.~Horava,
  ``Gluino condensation in strongly coupled heterotic string theory,''
  Phys.\ Rev.\  D {\bf 54}, 7561 (1996)
  [arXiv:hep-th/9608019].
}

\lref\dvg{
  R.~Dijkgraaf, S.~Gukov, V.~A.~Kazakov and C.~Vafa,
  ``Perturbative analysis of gauged matrix models,''
  Phys.\ Rev.\  D {\bf 68}, 045007 (2003)
  [arXiv:hep-th/0210238].
}

\lref\AV{
  M.~Aganagic and C.~Vafa,
  ``Mirror symmetry, D-branes and counting holomorphic discs,''
  [arXiv:hep-th/0012041].
}

\lref\vud{
  F.~Cachazo, B.~Fiol, K.~A.~Intriligator, S.~Katz and C.~Vafa,
  ``A geometric unification of dualities,''
  Nucl.\ Phys.\  B {\bf 628}, 3 (2002)
  [arXiv:hep-th/0110028].
}

\lref\civ{
  F.~Cachazo, K.~A.~Intriligator and C.~Vafa,
  ``A large $N$ duality via a geometric transition,''
  Nucl.\ Phys.\ B {\bf 603}, 3 (2001)
  [arXiv:hep-th/0103067].
} 

\lref\ckv{F.~Cachazo, S.~Katz and C.~Vafa,
  ``Geometric transitions and N = 1 quiver theories,''
  [arXiv:hep-th/0108120].
}

\lref\dvone{
  R.~Dijkgraaf and C.~Vafa,
  ``Matrix models, topological strings, and supersymmetric gauge theories,''
  Nucl.\ Phys.\  B {\bf 644}, 3 (2002)
  [arXiv:hep-th/0206255].
}

\lref\dvtwo{
  R.~Dijkgraaf and C.~Vafa,
  ``On geometry and matrix models,''
  Nucl.\ Phys.\  B {\bf 644}, 21 (2002)
  [arXiv:hep-th/0207106].
}

\lref\dvthree{
  R.~Dijkgraaf and C.~Vafa,
  ``A perturbative window into non-perturbative physics,''
  [arXiv:hep-th/0208048].
}

\lref\DGLVZ{
  R.~Dijkgraaf, M.~T.~Grisaru, C.~S.~Lam, C.~Vafa and D.~Zanon,
  ``Perturbative computation of glueball superpotentials,''
  Phys.\ Lett.\  B {\bf 573}, 138 (2003)
  [arXiv:hep-th/0211017].
}

\lref\coleman{
  S.~R.~Coleman,
  ``The Fate Of The False Vacuum. 1. Semiclassical Theory,''
  Phys.\ Rev.\ D {\bf 15}, 2929 (1977)
  [Erratum-ibid.\ D {\bf 16}, 1248 (1977)].
  C.~G.~.~Callan and S.~R.~Coleman,
  ``The Fate Of The False Vacuum. 2. First Quantum Corrections,''
  Phys.\ Rev.\ D {\bf 16}, 1762 (1977).
} 

\lref\gvw{
  S.~Gukov, C.~Vafa and E.~Witten,
  ``CFT's from Calabi-Yau four-folds,''
  Nucl.\ Phys.\  B {\bf 584}, 69 (2000)
  [Erratum-ibid.\  B {\bf 608}, 477 (2001)]
  [arXiv:hep-th/9906070].
}

\lref\BanksXH{
  T.~Banks, ``Landskepticism or why effective potentials don't count
  string models,'' 
  [arXiv:hep-th/0412129].  
}

\lref\MukhiDN{
  S.~Mukhi and N.~V.~Suryanarayana, ``A stable non-BPS configuration
  from intersecting branes and anti-branes,'' JHEP {\bf 0006}, 001 (2000)
  [arXiv:hep-th/0003219]. 
}

\lref\MukhiTE{
  S.~Mukhi, N.~V.~Suryanarayana and D.~Tong,
  ``Brane-anti-brane constructions,''
  JHEP {\bf 0003}, 015 (2000)
  [arXiv:hep-th/0001066].
}

\lref\silverstein{
  A.~Adams, J.~Polchinski and E.~Silverstein,
  ``Don't panic! Closed string tachyons in ALE space-times,''
  JHEP {\bf 0110}, 029 (2001)
  [arXiv:hep-th/0108075].
}

\lref\Klemm{
  S.~Chiantese, A.~Klemm and I.~Runkel,
  ``Higher order loop equations for A(r) and D(r) quiver matrix models,''
  JHEP {\bf 0403}, 033 (2004)
  [arXiv:hep-th/0311258].
}

\lref\Seki{
  S.~Seki,
  ``Comments on quiver gauge theories and matrix models,''
  Nucl.\ Phys.\  B {\bf 661}, 257 (2003)
  [arXiv:hep-th/0212079].
}

\lref\Eli{
  S.~Elitzur, A.~Giveon, D.~Kutasov, E.~Rabinovici and A.~Schwimmer,
  ``Brane dynamics and N = 1 supersymmetric gauge theory,''
  Nucl.\ Phys.\  B {\bf 505}, 202 (1997)
  [arXiv:hep-th/9704104].
}
\lref\Elit{
  S.~Elitzur, A.~Giveon and D.~Kutasov,
  ``Branes and N = 1 duality in string theory,''
  Phys.\ Lett.\  B {\bf 400}, 269 (1997)
  [arXiv:hep-th/9702014].
}
\lref\Feng{
  B.~Feng, A.~Hanany, Y.~H.~He and A.~M.~Uranga,
  ``Toric duality as Seiberg duality and brane diamonds,''
  JHEP {\bf 0112}, 035 (2001)
  [arXiv:hep-th/0109063].
}
\lref\Beasley{
  C.~E.~Beasley and M.~R.~Plesser,
  ``Toric duality is Seiberg duality,''
  JHEP {\bf 0112}, 001 (2001)
  [arXiv:hep-th/0109053].
}
\lref\Berenstein{
  D.~Berenstein and M.~R.~Douglas,
  ``Seiberg duality for quiver gauge theories,''
  arXiv:hep-th/0207027.
}
%


{\Title{\vbox{
\hbox{}
}}
{\vbox{
\centerline{Geometric Metastability, Quivers}
\vskip .1in
\centerline{and}
\vskip .1in 
\centerline{Holography}}}
\centerline{Mina Aganagic, Christopher Beem and Ben Freivogel}
\vskip .4in
}
\centerline{University of California, Berkeley, CA 94720}
\vskip .4in
\centerline{}
\vskip .4in
We use large $N$ duality to study brane/anti-brane configurations on a class of Calabi-Yau manifolds. With only branes present, the Calabi-Yau manifolds in question give rise to ${\cal N}=2$ ADE quiver theories deformed by superpotential terms. We show that the large $N$ duality conjecture of \Aganagic\ reproduces correctly the known qualitative features of the brane/anti-brane physics. In the supersymmetric case, the gauge theories have Seiberg dualities, which are represented as flops in the geometry. Moreover, the holographic dual geometry encodes the whole RG flow of the gauge theory. In the non-supersymmetric case, the large $N$ duality predicts that the brane/anti-brane theories also enjoy such dualities, and allows one to pick out the good description at a given energy scale.

\vfill
\eject
\newsec{Introduction}

Geometric transitions have proven to be a powerful means of studying
the dynamics of supersymmetric D-branes. String theory relates these
transitions to large $N$ dualities, where before the transition, at 
small 't Hooft coupling, one has D-branes wrapping cycles in the 
geometry, and after the transition, at large 't Hooft coupling, the 
system is represented by a different geometry, with branes replaced 
by fluxes. The AdS/CFT correspondence can be thought of in this way. 
Geometric transitions are particularly powerful when the D-branes in 
question wrap cycles in a Calabi-Yau manifold. Then, the topological 
string can be used to study the dual geometry exactly to all orders 
in the 't Hooft coupling.  In \Aganagic\ it was conjectured that 
topological strings and large $N$ dualities can also be used to study
non-supersymmetric, metastable configurations of branes in Calabi-Yau
manifolds, that confine at low energies.  This conjecture was considered in greater detail in \refs{\Seo,\Heckman}. String theory realizations of metastable, supersymmetry breaking vacua have appeared in \refs{\KachruVJ\VafaWI\KachruGS\KachruAW\ArgurioNY\VerlindeBC\ArgurioQK\MurthyQM\MarsanoFE-\douglas}.  The gauge theoretic mechanism of \ISS\ has further been explored in string theory in \refs{\OoguriPJ\OoguriBG\FrancoHT\Bena\Ahn\TatarDM\GiveonFK-\TatarZA}.

In this paper we study D5 brane/anti-D5 brane systems in IIB on non-compact, 
Calabi-Yau manifolds that are ADE type ALE space fibrations 
over a plane. These generalize the case of the $A_1$ ALE space studied 
in detail in \refs{\Aganagic, \Seo, \Heckman}. The ALE space is fibered over the 
complex plane in such a way that at isolated points, the 2-cycles 
inherited from the ALE space have minimal area. These minimal 2-cycles 
are associated to positive roots of the corresponding ADE Lie algebra. 
Wrapping these with branes and anti-branes is equivalent to considering 
only branes, but allowing both positive and {\it negative} roots to appear, 
corresponding to two different orientations of the $S^2$'s. The system 
can be metastable since the branes wrap {\it isolated} minimal 2-cycles, 
and the cost in energy for the branes to move, due to the tensions of the 
branes, can overwhelm the Coulomb/gravitational attraction between them.

The geometries in question have geometric transitions in which the
sizes of the minimal $S^2$'s go to zero, and the singularities are
resolved instead by finite sized $S^3$'s. The conjecture of \Aganagic\
is that at large $N$, the $S^2$'s disappear along with the branes and
anti-branes and are replaced by $S^3$'s with positive and negative
fluxes, the sign depending on the charge of the replaced branes.  As
in the supersymmetric case (see \refs{\civ,\ckv,\vud}), the dual
gravity theory has ${\cal N}=2$ supersymmetry softly broken to ${\cal
N}=1$ by the fluxes. The only difference is that now some of the
fluxes are negative. On-shell, the positive and the negative fluxes
preserve different halves of the original supersymmetry, and with both
present, the ${\cal N}=2$ supersymmetry is completely broken in the
vacuum (see \HoravaVS\ for discussion of a similar supersymmetry
breaking mechanism and its phenomenological features in the context of heterotic M-theory). The topological string computes not only
the superpotential, but also the K\"ahler potential.\foot{While the
superpotential is exact, the K\"ahler potential is not.  Corrections to the K\"ahler potential coming from warping, present 
when the Calabi-Yau is compact, have been investigated in
\douglas .} We show that the Calabi-Yau's with fluxes obtained in
this way are indeed metastable, as expected by holography. In
particular, for widely separated branes, the supersymmetry breaking
can be made arbitrarily weak.\foot{ The natural measure of
supersymmetry breaking in this case is the mass splitting between the
bosons and their superpartners.  For a compact Calabi-Yau, the scale
of supersymmetry breaking is set by the mass of the gravitino, which is of the order of the cosmological constant. In our case, gravity is not
dynamical, and the mass splittings of the dynamical fields are tunable
\Aganagic .} In fact, we can use the gravity dual to learn about the
physics of branes and anti-branes. We find that at one-loop, the
interaction between the branes depends on the topological data of the
Calabi-Yau in a simple way. Namely, for every brane/anti-brane pair,
so for every positive root $e_+$ and negative root $e_-$, we find that
the branes and the anti-branes attract if the inner product
$$e_+ \cdot e_-$$
is {\it positive}.  They repel if it is negative, and do not interact at
all if it is zero. In the $A_k$ type ALE spaces, this result is already 
known from the direct open string computation \refs{\MukhiDN,\MukhiTE}, 
so this is a simple but nice test of the conjecture for these geometries. 
Moreover, we show that certain aspects of these systems are universal. 
We find that generically, just like in \Seo, metastability is lost 
when the 't Hooft coupling becomes sufficiently large. Moreover, once 
stability is lost, the system appears to roll down toward a vacuum in 
which domain walls interpolating between different values of the fluxes 
become light.  We also present some special cases where the 
non-supersymmetric brane/anti-brane systems are exactly stable. In these 
cases, there are no supersymmetric vacua to which the system can decay.

When all the branes are D5 branes and supersymmetry is preserved, the
low energy theory geometrically realizes \refs{\ckv,\vud} a 4d ${\cal
N}=2$ supersymmetric quiver gauge theory with a superpotential for the
world-volume adjoints which breaks ${\cal N}=2$ to ${\cal N}=1$.
These theories are known to have Seiberg-like dualities \seiberg\ in which the dual theories flow to the same IR fixed point, and where different
descriptions are more weakly coupled, and hence preferred, at different
energy scales. The Seiberg dualities are realized in the geometry in a
beautiful way \vud . The ADE fibered Calabi-Yau geometries used to
engineer the gauge theories have {\it intrinsic} ambiguities in how one
resolves the singularities by blowing up $S^2$'s. The different
possible resolutions are related by flops that shrink some 2-cycles,
and blow up others. The flops act non-trivially on the brane charges,
and hence on the ranks of the gauge groups. The flop of a 2-cycle
$S^2_{i_0}$ corresponds to a Weyl reflection about the corresponding
root of the Lie algebra. On the simple roots $e_i$, this acts by
$$
S^2_i \rightarrow {\tilde S}^2_i = S^2_i - (e_i \cdot e_{i_0})\, S^2_{i_0}.
$$
Brane charge conservation then implies that the net brane charges transform satisfying
\eqn\first{
\sum_{i} N_i \,S^2_i = \sum_{i} {\tilde N}_i\, {\tilde S}^2_i.
}
Moreover, from the dual gravity solution one can reconstruct the whole
RG flow of the gauge theory.  The sizes of the wrapped 2-cycles
encode the gauge couplings, and one can read off how these vary over
the geometry, and correspondingly, what is the weakly coupled
description at a given scale.  Near the $S^3$'s, close to where the
branes were prior to the transition, corresponds to long distances in
the gauge theory.  There, the $S^2$'s have shrunken, corresponding to
the fact that in the deep IR the gauge theories confine.  As one goes
to higher energies, the gauge couplings may simply become weaker, and
the corresponding $S^2$'s larger, in which case the same theory will
describe physics at all energy scales. Sometimes, however, some of the
gauge couplings grow stronger, and the areas of the $S^2$'s eventually
become negative. Then, to keep the couplings positive, the geometry
must undergo flop transitions.\foot{It is important, and one can verify
this, that this happens in a completely {\it smooth} way in the geometry,
as the gauge coupling going to infinity corresponds to zero K\"ahler volume
of the 2-cycle, while the physical size of the 2-cycle is finite
everywhere away from the $S^3$'s.}  This rearranges the brane charges
and corresponds to replacing the original description at low energies by a different one at high energies.  Moreover, the flops of the
$S^2$'s were found to coincide exactly with Seiberg dualities of the
supersymmetric gauge theories.
  
In the non-supersymmetric case we do not have gauge theory
predictions to guide us. However, the string theory still has
intrinsic ambiguities in how the singularities are resolved.
This is {\it exactly} the same as in the supersymmetric case, except that
now not all $N_i$'s in \first\ need be positive. 
Moreover, we can use holography to 
follow the varying sizes of 2-cycles over the
geometry, and find that indeed in some cases 
they can undergo flops in going from the IR to the UV. 
When this happens, descriptions in terms of different brane/anti-brane 
configurations are more natural at different energy scales, and one can smoothly 
interpolate between them. This is to be contrasted with, say, the $A_1$ 
case, where regardless of whether one considers just branes or branes and anti-branes, it is only one description that is ever really 
weakly coupled, and the fact that another exists is purely formal.

The paper is organized as follows. In section 2 we introduce the
metastable D5 brane/anti-D5 brane configurations, focusing on $A_k$
singularities, and review the conjecture of \Aganagic\ applied to this
setting.  In section 3 we study in detail the $A_2$ case with a
quadratic superpotential.  In section 4 we consider general ADE type
geometries.  In section 5 we discuss Seiberg-like dualities of these
theories. In section 6 we study a very simple, exactly solvable
case. In appendices A and B, we present the matrix model computation
of the prepotential for $A_2$ ALE space fibration, as well as the
direct computation from the geometry. To our knowledge, these
computations have not been done before, and the agreement provides a direct
check of the Dijkgraaf-Vafa conjecture for these geometries. Moreover,
our methods extend easily to the other $A_n$ cases. In appendix C, we
collect some formulas useful in studying the metastability of our
solutions in section 3.

\newsec{Quiver Branes and Anti-branes}

Consider a Calabi-Yau which is an $A_{k}$ type ALE space, 
\eqn\fibr{
x^2+y^2+\prod_{i=1}^{k+1}(z-z_i(t))= 0,
}
fibered over the $t$ plane. Here, $z_i(t)$ are polynomials in $t$.  Viewed as a family of ALE spaces parameterized by $t$, there are $k$ vanishing 
2-cycles, 
\eqn\hom{S^2_{i}, \qquad i=1,\ldots, k}
that deform the the singularities of \fibr. In the fiber over each point $t$ in the base, the 2-cycle in the class $S^2_i$ has holomorphic area given by
\eqn\twof{
\int_{S^2_{i,t}} \omega^{2,0} = z_{i}(t)-z_{i+1}(t).
}
where $\omega^{2,0}$ is the reduction of the holomorphic three-form $\Omega$ on the fiber.  The only singularities are at points where $x=y=0$ and
\eqn\crit{
z_i(t)=z_j(t),\qquad i\neq j
}
for some $i$ and $j$.  At these points, the area of one of the 2-cycles inherited from the ALE space goes to zero. 

These singularities can be smoothed out by blowing up the 2-cycles, i.e., by changing the K\"ahler structure of the Calabi-Yau to give them all non-vanishing area.\foot{As we will review later, the blowup is not unique, as not all the K\"ahler areas of the cycles in \hom\ need to be positive for the space to be smooth. Instead, there are different possible blowups which differ by flops.}  The homology classes of the vanishing cycles \crit\ then correspond to positive roots of the $A_k$ Lie algebra (see e.g. \ckv ).\foot{The negative roots correspond to 2-cycles of the opposite orientation.} In this case, the $k$ simple, positive roots $e_i$ correspond to the generators of the second homology group. These are the classes of the $S^2_i$ mentioned above which resolve the singularities where $z_i(t)=z_{i+1}(t)$. We denote the complexified K\"ahler areas of the simple roots by
$$
r_i= \int_{S^2_{i}} k+iB^{NS},
$$ 
where $k$ is the K\"ahler form. In most of our applications, we'll take the real part of $r_i$ to vanish. The string theory background is
non-singular as long as the imaginary parts do not also vanish. They are positive, per definition, since we have taken the $S^2_i$ to correspond to positive roots.  In classical geometry, the $r_i$ are independent of $t$. Quantum mechanically, in the presence of branes, one finds that they are not. 

There are also positive, non-simple roots
$e_{I}=\sum_{i=j}^{l}e_i$, for $l>j$
where $z_{l+1}(t)=z_{j}(t)$. The 2-cycle that resolves the singularity is given by
$$
S^2_{I}=\sum_{i=j}^l S^2_{i}
$$
in homology.  Its complexified K\"ahler area is given as a sum of K\"ahler areas of simple roots
$$
r_{I}=\sum_{i=j}^l r_i.
$$
The total area $A(t)$ of a 2-cycle $S^2_I$ at a fixed $t$ receives contributions from both K\"ahler and holomorphic areas:
\eqn\area{
A_{I}(t) = \sqrt{|r_I|^2+ |W_I'(t)|^2}.
}
The functions $W'_{I}$ capture the holomorphic volumes of 2-cycles, and are related to the geometry by
$$
W_{I}(t) = \sum_{i=j}^k W_{i}(t),
$$
\eqn\supp{
W_i(t) = \int (z_{i}(t)-z_{i+1}(t)) dt.
}
These will reappear as superpotentials in matrix models which govern the open and closed topological string theory on these geometries.

For each positive root $I$ there may be more than one solution to \crit .  We will label these with an additional index $p$ when denoting the corresponding 2-cycles, $S^2_{I,p}$. For each solution there is an isolated, minimal area $S^2$, but they are all in the same homology class, labeled by the root. They have minimal area because \area\ is minimized at those points in the $t$ plane where $W'_{I}(t)$ vanishes.  These, in turn, correspond to solutions of \crit .

We will consider wrapping branes in the homology class
$$
\sum_{I,p} M_{I,p} \, S^2_I,
$$
with $I$ running over {\it all} positive roots, and $p$ over the corresponding critical points.  We get branes or anti-branes on $S^2_{I,p}$ depending on whether the charge $M_{I,p}$ is positive or negative.\foot{We could have instead declared all the $M_{I,p}$ to be positive, and summed instead over positive and negative roots.}  We will study what happens when we wrap branes on some of the minimal $S^2$'s and anti-branes on others. 

The brane/anti-brane system is not supersymmetric.  If we had branes wrapping all of the $S^2$'s, they would have each preserved the same half of the original $\CN=2$ supersymmetry.  However, with some of the branes replaced by anti-branes, some stacks preserve the opposite half of the original supersymmetry, and so globally, supersymmetry is completely broken. The system can still be metastable. As in flat space, there can be attractive Coulomb/gravitational forces between the branes and the anti-branes.  For them to annihilate, however, they have to leave the minimal 2-cycles that they wrap. In doing so, the area of the wrapped 2-cycle increases, as can be seen from \area, and this costs energy due to the tension of the branes.  At sufficiently weak coupling, the Coulomb and gravitational interactions should be negligible compared to the tension forces -- the former are a one-loop effect in the open string theory, while the latter are present already at tree-level -- so the system should indeed be metastable.  For this to be possible, it is crucial that the parameters of the background, {\it i.e.} the K\"ahler moduli $r_i$ and the complex structure moduli that enter into the $W_i(t)$, are all non-normalizable, and so can be tuned at will.

While this theory is hard to study directly in the open string language, it was conjectured in \Aganagic\ to have a holographic dual which gives an excellent description when the number of branes is large.

\subsec{Supersymmetric Large $N$ Duality}

Here we review the case where only branes are wrapped on the minimal $S^2$'s, and so supersymmetry is preserved.  Denoting the net brane charge in the class  $S^2_i$ by $N_i$, this geometrically engineers an ${\cal N}=2$ supersymmetric $\prod_{i=1}^k U(N_i)$ quiver gauge theory in four dimensions, deformed to ${\cal N}=1$ by the presence of a superpotential. The corresponding quiver diagram is the same as the
Dynkin diagram of the $A_k$ Lie algebra.  The $k$ nodes correspond to the $k$ gauge groups, and the links between them to bifundamental
hypermultiplets coming from the lowest lying string modes at the intersections of the $S^2$'s in the ALE space.  The superpotential for the adjoint valued chiral field $\Phi_{i}$, which breaks the supersymmetry to ${\cal N}=1$, is
$$
W_i(\Phi_{i}), \qquad i=1,\ldots k
$$
where $W_i(t)$ is given in \supp. The chiral field $\Phi_i$ describes the position of the branes on the $t$ plane. As shown in \ckv , the gauge theory has many supersymmetric vacua, corresponding to all possible ways of distributing the branes on the $S^2$'s,
$$ 
\sum_{i=1}^k N_i \, S^2_i = \sum_{I,p} M_{p,I}\, S^2_I,
$$
where $I$ labels the positive roots and $p$ the critical points associated with a given root.  This breaks the gauge symmetry as
\eqn\gauge{
\prod_i U(N_i) \rightarrow \prod_{p,I} U(M_{p,I}).
}
At low energies the branes are isolated and the theory is a pure ${\cal N}=1$ gauge theory with gauge group \gauge .  The $SU(M_{I,p})$ subgroups of the $U(M_{I,p})$ gauge groups experience confinement and gaugino condensation.

This theory has a holographic, large $N$ dual where branes are replaced by fluxes.  The large $N$ duality is a geometric transition
which replaces \fibr\ with a dual geometry
\eqn\fibrdual{
x^2+y^2+\prod_{i=1}^{k+1}(z-z_i(t))= f_{r-1}(t) z^{k-1} + 
f_{2r-1}(t) z^{k-2} +\ldots + f_{kr-1}(t) ,
}
where $f_{n}(t)$ are polynomials of degree $n$, with $r$ being the highest of the degrees of $z_i(t)$.  The geometric transition replaces each of the $S^2_{I,p}$'s by a three-sphere, which will be denoted $\CA_{I,p}$, with $M_{I,p}$ units of Ramond-Ramond flux through it,
$$
\int_{\CA_{I,p}} H^{RR} + \tau H^{NS} = M_{I,p}.
$$
In addition, there is flux through the non-compact dual cycles $\CB_{I,p}$,
$$
\int_{\CB_{I,p}} H^{RR} + \tau H^{NS} = -\alpha_I,
$$
where $\tau$ is the IIB axion-dilaton $\tau=a+{i\over g_s}$. These cycles arise by fibering $S^2_{I,p}$ over the $t$ plane, with the 2-cycles vanishing at the branch cuts where the $S^3$'s open up. The nonzero $H$ flux through the $B$-type cycles means that 
$$
\int_{S^2_{I,p}} B^{RR}+ \tau B^{NS}
$$
varies over the $t$ plane. In the gauge theory, this combination determines the complexified gauge coupling. Since 
$$
{4 \pi\over g_i^2} = {1 \over g_s} \int_{S^2_i} B^{NS}, \qquad
{\theta_i\over 2\pi}=
\int_{S^2_i} B^{RR} + a B^{NS}, 
$$
one naturally identifies $\alpha_i$ with the gauge coupling of the $U(N_i)$, ${\cal N}=2$ theory at a high scale\foot{For the large $N$ dual to be an honest Calabi-Yau, as opposed to a generalized one, we will work with $\int_{S^2_i} k=0$.}   
$$
\alpha_i = -{\theta_i \over 2\pi} - {4\pi i\over g_i^2}. 
$$
For each positive root $I$, we then define $\alpha_{I}$ as
$$
\alpha_{I} =\sum_{i=j}^{k} \alpha_i
$$

Turning on fluxes gives rise to an effective superpotential \gvw\
$$
\CW_{\rm eff} = \int_{CY} (H^{RR}+\tau H^{NS}) \wedge \Omega.
$$
Using the special geometry relations
$$
\int_{A_{I,p}} \Omega = S_{I,p}, \qquad 
\int_{B_{I,p}} \Omega = \del_{S_{I,p}} {\cal F}_0, 
$$
the effective superpotential can be written as 
\eqn\effsupp{
\CW_{\rm eff} = \sum_{I,p} \alpha_I\; S_{I,p} + M_{I,p}\; {\del_{S_{I,p}}} {\cal F}_0.
}
Here, $S_{I,p}$ gets identified with the value of the gaugino bilinear of the $U(M_{I,p})$ gauge group factor on the open string side.  The effective superpotential \effsupp\ can be computed directly in the gauge theory. Alternatively, it can be shown \refs{\dvthree,\DGLVZ} that the relevant computation reduces to computing planar diagrams in a gauged matrix model given by the zero-dimensional path integral
$$ 
{1\over \prod_{i=1}^k \vol\; U(N_i)}\;\; \int\prod_{i=1}^k d
\Phi_i \, dQ_{i,i+1} d{Q}_{i+1,i}\;\;\;
\exp\Bigl(-{1\over g_s }{\rm Tr} \CW(\Phi, Q)\Bigr)
$$
where
$$
{\rm Tr}\, \CW(\Phi, Q)= 
\sum_{i=1}^r {\rm Tr}\, W(\Phi_i)+ 
{\rm Tr}\,( Q_{i+1,i} \Phi_i {Q}_{i,i+1} - {Q}_{i,i+1} \Phi_{i+1} Q_{i+1,i}).
$$   

The critical points of the matrix model superpotential correspond to the supersymmetric vacua of the gauge theory. 
The prepotential ${\cal F}_0(S_{I,p})$ that enters the superpotential \effsupp\ is the planar free energy of the matrix model 
\refs{\dvthree,\dvone,\dvtwo,\dvg}, expanded about a critical point where the gauge group is broken 
as in \gauge .
More precisely, we have
$$
2\pi i {\cal F}_0(S) = {\cal F}_{0}^{\,np}(S) + \sum_{\{h_a\}} {\cal F}_{0,\{h_a\}}\prod_{a} S_a^{h_a}
$$
where ${\cal F}_{0,\{h_a\}} \prod_a (M_a g_s)^{h_a}$ 
is the contribution to the planar free
energy coming from diagrams with $h_a$ boundaries carrying the index of
the $U(M_a)$ factor of the unbroken gauge group. Here
$a$ represents a pair of indices,
$$
a=(I,p),
$$ 
and we've denoted 
$S_{a} = M_{a} g_s$. 
The ``non-perturbative'' contribution,
${\cal F}_{0}^{\,np}(S)$, to the matrix model amplitude comes from the
volume of the gauge group \gauge\ that is {\it unbroken} in the vacuum at
hand \refs{\dvone,\dvg}, and is the prepotential of the leading order
conifold singularity corresponding to the shrinking $S^3$, which is
universal. 
We will explain how to compute the matrix integrals in appendix $A$.  
The supersymmetric vacua of the theory are then given by
the critical points of the superpotential $\CW_{\rm eff}$,
$$
\del_{S_{a}} \CW_{\rm eff}=0.
$$
\subsec{Non-Supersymmetric Large $N$ Duality}

Now consider replacing some of the branes with anti-branes while keeping the background fixed.  The charge of the branes, as measured at infinity, is computed by the RR flux through the $S^3$ that surrounds the branes.  In the large $N$ dual geometry, the $S^3$ surrounding the wrapped $S^2_{I,p}$ is just the cycle $\CA_{I,p}$. Replacing the branes with anti-branes on some of the $S^2$'s then has the effect of changing the signs of the  corresponding $M_{I,p}$'s. Moreover, supersymmetry is now broken, so the vacua of the theory will appear as critical points of the physical potential
\eqn\pot{
V = G^{S_{a}{\bar S}_{b}} \; \del_{S_{a}} \CW_{\rm eff} 
\;{\overline{\del_{{S}_{b}} {\CW_{\rm eff}}}} +V_0.
}
The superpotential $\CW_{\rm eff}$ is still given by \effsupp , and $G$ is the K\"ahler metric of the ${\cal N}=2$ theory,
$$
G_{a{\bar b}} = {\rm Im}(\tau)_{{a} {\bar b}}
$$
where
$$
\tau_{ab} = \del_{S_a} \del_{S_b} {\cal F}_0
$$
and $a,b$ stand for pairs of indices $(I,p)$.  In the absence of gravity, we are free to add a constant, $V_0$, to the potential,\foot{This simply adds a constant to the Lagrangian, having nothing to do with supersymmetry, or its breaking.} which we'll take to be
\eqn\zero{
V_0 = \sum_{I,p} {M_{I,p} \over g_{I}^2}.
}
{\it A priori}, $V_0$ can be either positive or negative, 
depending on the charges. However, we'll see that in all the vacua where the theory is weakly coupled, the leading contribution to the effective potential at the critical point will turn out to be just the tensions of all the branes, which is strictly positive.

\newsec{A Simple Example}

We now specialize to an $A_2$ quiver theory with quadratic superpotential. 
The geometry which engineers this theory is given by \fibr, with
$$
z_1(t) =-m_1(t-a_1), \qquad z_2(t) = 0,\qquad z_3(t)=m_2(t-a_2).
$$
There are three singular critical points \crit\ (assuming generic $m_i$) corresponding to
$$
t = a_i, \qquad i=1,2,3
$$
where $a_3 = (m_1 a_1+m_2 a_2)/(m_1+m_2)$.
Blowing up to recover a smooth Calabi-Yau, the singular points are replaced by three positive area $S^2$'s,
$$
S^2_1, \; S^2_2, \; S^2_3
$$
with one homological relation among them,
\eqn\homtwo{
S^2_{3}=S^2_1+S^2_2.
}
$S^{2}_{1,2}$ then correspond to the two simple roots of the $A_2$ Lie algebra, $e_{1,2}$, and $S^2_3$ is the one non-simple positive root, $e_1+e_2$. Now consider wrapping branes on the three minimal 2-cycles so that the total wrapped cycle $C$ is given by
$$
C=M_1 \, S^2_1+M_2 \, S^2_2+M_3 \,S^2_3.
$$
If some, but not all, of the $M_I$ are negative, supersymmetry is broken. As was explained in the previous section, as long as the branes are widely separated, this system should be perturbatively stable.

\bigskip
\centerline{\epsfxsize 3.3truein\epsfbox{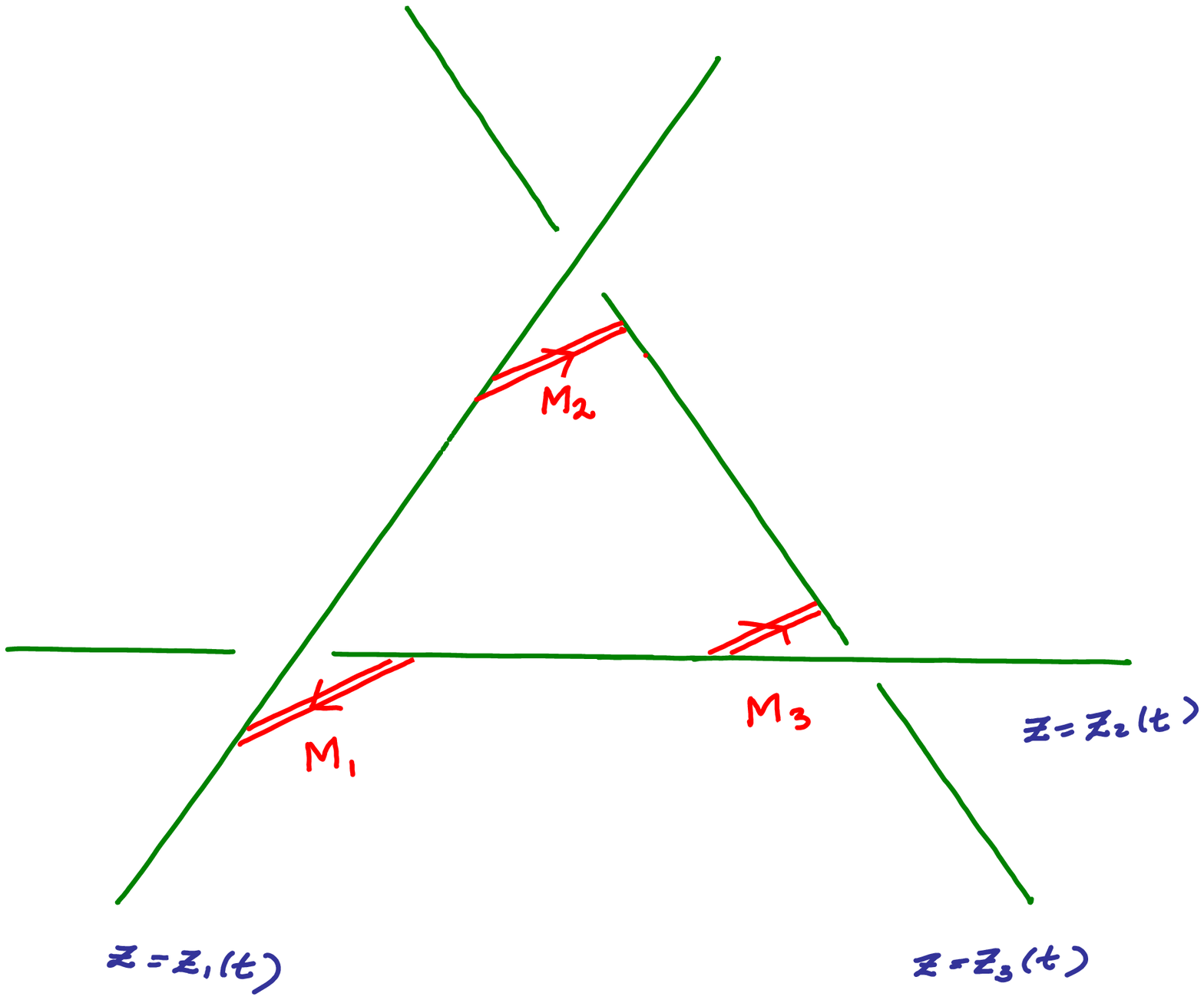}}
\noindent{\ninepoint
\baselineskip=2pt {\bf Fig. 1.} {The figure corresponds to the $A_2$ singularity in the $z$-$t$ plane with quadratic ``superpotential''. There are three conifold singularities at $z_i=z_j$ which can be blown up by three $S^2$'s, spanning two homology classes. Wrapping $M_1$ anti-D5 branes on $S^2_1$ and $M_{2,3}$ D5 branes on $S^2_{2,3}$, we can engineer a metastable vacuum. The orientations of the branes are indicated by arrows. }}
\bigskip
Non-perturbatively, we expect the branes to be able to tunnel to a lower energy state. The minimum energy configuration that this system
can achieve depends on the net brane charges in the homology classes $S^2_1$ and $S^2_2$, given by $N_1=M_1+M_3$
and $N_2 = M_2+M_3$. When $N_1$ and $N_2$ have the same sign, the system can tunnel to a supersymmetric vacuum with new charges
$$
M_I\rightarrow M_I'
$$
where all the $M_I'$ share the same sign, and the net charges $N_1=M_1'+M_3'$ and $N_2=M_2'+M_3'$ are unchanged. All the supersymmetric vacua are degenerate in energy, but for the metastable, non-supersymmetric vacua, the decay rates will depend on the $M_I'$.
Alternatively, if one of the $N_{1,2}$ is positive and the other is negative, the lowest energy configuration is necessarily not supersymmetric.  In this way we get a stable, non-supersymmetric state which has nowhere to which it can decay.

In the remainder of this section, we will study these systems using the large $N$ dual geometry with fluxes.

\subsec{The Large $N$ Dual} 

The large $N$ dual geometry in this case is given by
\eqn\cs{
x^2+y^2+z(z-m_1(t-a_1))(z+m_2(t-a_2)) = c z+d t+ e.
}
The three $S^2$'s at the critical points have been replaced by three $S^3$'s, $\CA_{I}$, whose sizes are related to the coefficients $c,d,e$ 
above.  There are also three non-compact, dual 3-cycles $\CB_I$.  The geometry of the Calabi-Yau is closely related to the geometry of the
Riemann surface obtained by setting $x=y=0$ in \fibrdual.  The Riemann surface can be viewed as a triple cover of the $t$ plane, by writing \cs\ as
$$
0 = (z-z'_1(t))(z-z'_2(t))(z-z'_3(t))
$$
where $z'_i(t)$ correspond to the $z_i(t)$ which are deformed in going from \fibr\ to \cs\ . In particular, the holomorphic three-form $\Omega$ of
the Calabi-Yau manifold descends to a 1-form on the Riemann surface, as can be seen by writing
$$
\Omega = \omega^{2,0} \wedge dt
$$
and integrating $\omega^{2,0}$ over the $S^2$ fibers, as in \twof .  The $A$ and $B$ cycles then project to 1-cycles on the Riemann surface.
The three sheets are glued together over branch cuts which open up at $t=a_I$.  We have
$$
S_I= {1\over 2 \pi i} \int_{{a_I}^-}^{{a_I}^+}(z'_{J}(t)-z'_K(t))\; dt, 
\qquad\;
\del_{S_I} {\cal F}_0= {1\over 2 \pi i}\int_{{a_I}^+}^{\Lambda_0} \;(z'_J(t)-z'_K(t)) \;dt
$$ 
for cyclic permutations of distinct $I,J$ and $K$. This allows one to
compute the prepotential ${\cal F}_0$ by direct
integration (see appendix B).
Alternatively, by the conjecture of \dvone, the same
prepotential can be computed from the corresponding matrix model. The
gauge fixing of the 
matrix model is somewhat involved, and we have
relegated it to appendix A, but the end result is very simple.  The field content consists of:
\item{{a.}}{Three sets of adjoints $\Phi_{ii}$ of $U(M_i)$, which describe the fluctuations of the branes around the three $S^2$'s.}
\vskip 0.1cm
\item{{b.}}{A pair of bifundamental matter fields $Q_{12}$, ${\tilde Q}_{21}$, coming from the 12 strings. 
}
\vskip 0.1cm
\item{{c.}}{Anti-commuting bosonic ghosts, 
$B_{13}$, ${C}_{31}$ and $B_{32}$, ${C}_{23}$, representing the 23 and 31 strings.
} 

\noindent Note that physical bifundamental matter from $S^2$'s with positive intersection corresponds to commuting 
bosonic bifundamentals in the matrix model, whereas $W$ bosons between $S^2$'s with negative intersection in the physical theory
correspond to bosonic ghosts, similarly to what happened in \dvg. 

The effective superpotential for these fields is
$$
{\eqalign{
\CW_{\rm eff} =&  {1\over 2}m_1\, \Tr \Phi_{11}^2+{1\over 2}m_2 \,\Tr \Phi_{22}^2+
{1\over 2} m_3 \, \Tr \Phi_{33}^2\cr
+&\,a_{12}\, \Tr {Q}_{12}{\tilde Q}_{21}
+\,
a_{23}\, \Tr {B}_{32}{C}_{23}
+\,
a_{31} \,\Tr{B}_{13}{C}_{31}\cr
+&\,
\Tr
({B}_{32} \Phi_{22}{C}_{23}-C_{23}\Phi_{33}B_{32})
+\,
\Tr
({B}_{13} \Phi_{33}{C}_{31}-C_{31}\Phi_{11}B_{13})\cr
+&\,
\Tr
( {\tilde Q}_{21} \Phi_{11}{Q}_{12}-Q_{12}\Phi_{22}{\tilde Q}_{21})
}}
$$
where $a_{ij} = a_i-a_j$. 
From this we can read off the propagators
$$
\langle \Phi_{ii} \Phi_{ii} \rangle = {1\over m_i},\qquad
\langle Q_{12} {\tilde Q}_{21} \rangle = {1\over a_{12}}
$$
and 
$$
\langle B_{23}C_{32} \rangle = -{1\over a_{23}}, \qquad
\langle B_{31}C_{13} \rangle = -{1\over a_{31}}, 
$$
as well as the vertices. 
\bigskip
\bigskip
\centerline{\epsfxsize 4.0truein\epsfbox{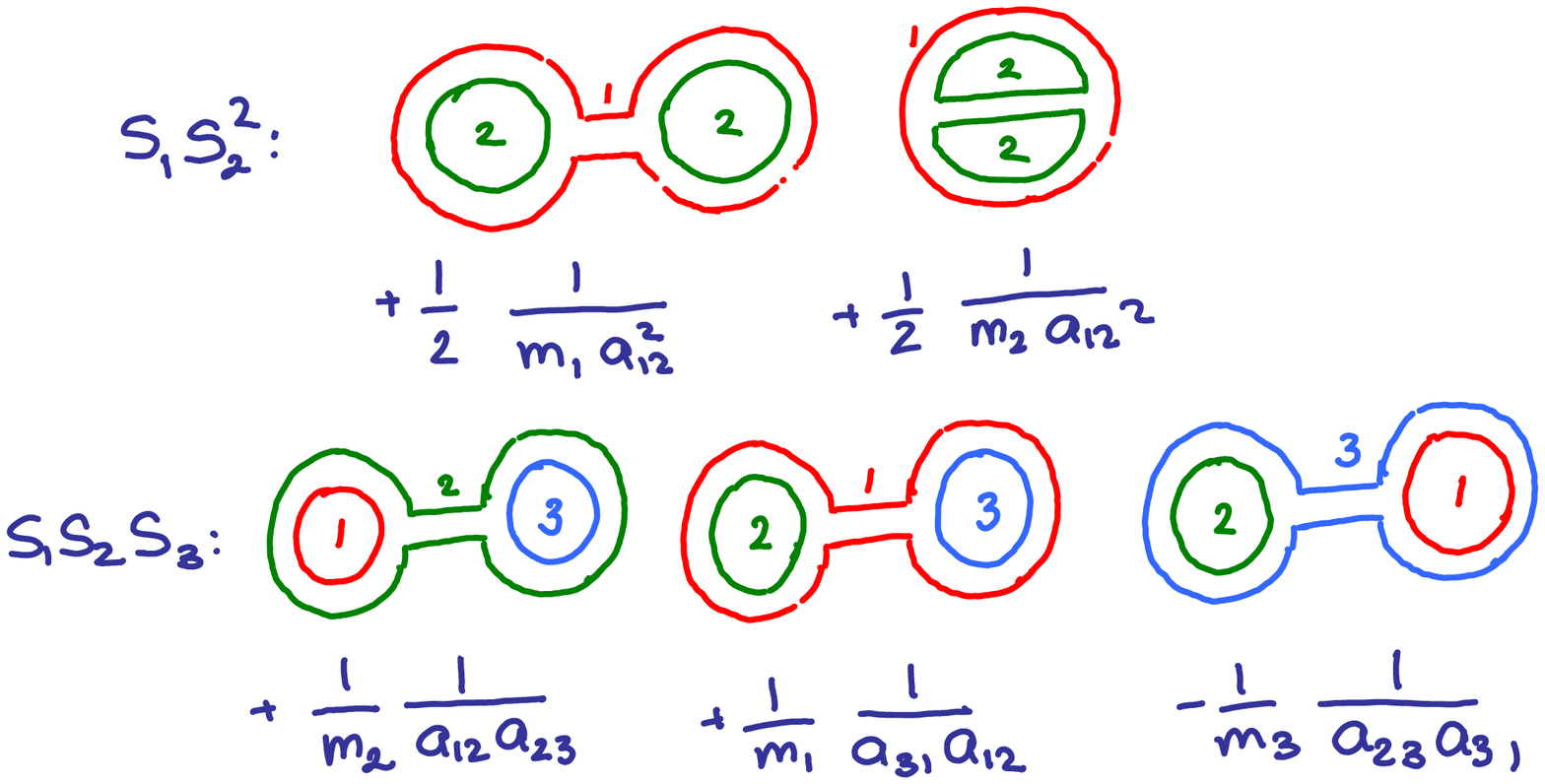}}
\noindent{\ninepoint
\baselineskip=2pt {\bf Fig. 2.} 
{
Some of the two-loop Feynman graphs of the matrix model path 
integral, which are computing the prepotential ${\cal F}_{0}$. 
The path integral is expanded about a vacuum corresponding to distributing branes on the three nodes. Here, the boundaries on node one are colored red, on node two are green and on node three are blue. 
}}
\bigskip
Keeping only those contractions of color indices that correspond to planar diagrams, and carefully keeping track of the signs associated with fermion loops, we find:
$$\eqalign{
2 \pi i {\cal F}_0(S_i)=& 
\;{1\over 2}\, S_1^2\,(\log({S_1\over m_1\Lambda_0^2}) - {3\over 2}\,)+
\;{1\over 2}\, S_2^2\,(\log({S_2\over m_2\Lambda_0^2}) - {3\over 2}\,)+
\;{1\over 2}\, S_3^2\,(\log({S_3\over m_3\Lambda_0^2}) - {3\over 2}\,)\cr
&
\; - \log({a_{12}\over \Lambda_0})\, S_1S_2 + 
\log({a_{31}\over \Lambda_0}) \, S_1 S_3+\log({a_{23}\over \Lambda_0})
\, S_2 S_3\cr
&
+{1\over 2 \Delta^3}(S_1^2 S_2 + S_2^2 S_1+S_3^2 S_1+
S_3^2 S_2 - S_1^2 S_3 - S_2^2 S_3- 6 S_1S_2S_3)+\ldots}
$$
where
\eqn\DD{
\Delta^3 = {m_1 m_2\over m_3}a_{12}^2,\qquad\qquad m_3=m_1+m_2.
}
The terms quadratic in the $S_i$'s correspond to one-loop terms in the matrix model, the cubic terms to two-loop terms, and so on.
The fact that the matrix model result agrees with 
the direct computation from the geometry is a nice direct check of the Dijkgraaf-Vafa conjecture for quiver theories.  The large $N$ limit of quiver matrix models was previously studied using large $N$ saddle point techniques in \refs{\ckv,\vud,\Klemm,\Seki}.

Consider now the critical points of the potential \pot,
$$
\del_{S_I} V = 0.
$$
The full potential is very complicated, but at weak 't Hooft coupling (we will show this is consistent {\it a posteriori}) it should be sufficient to keep only the leading terms in the expansion of ${\cal F}_0$ in powers of ${S/\Delta^3}$. These correspond to keeping only the one-loop terms in the matrix model.  In this approximation, the physical vacua of the potential \pot\ correspond to solutions of 
\eqn\solone{
\alpha_I + 
\sum_{M_J>0}\tau_{IJ}M^{J} +
\sum_{M_J<0}{\bar \tau}_{IJ}M^{J}  
=0.
}
To be more precise, there {\it are} more solutions with other sign choices for $\pm M_J$, but only {\it this} choice leads to ${\rm Im}(\tau)$ being positive definite.  Since ${\rm Im}(\tau)$ is also the metric on the moduli space, only this solution is physical.

Depending on how we choose to distribute the branes, there are two distinct  classes of non-supersymmetric vacua which can be constructed in this way.  We will discuss both of them presently.

\subsec{$M_1<0$, $M_{2,3}>0$}

In this case, the critical points of the potential correspond to
$$
\eqalign{
&{\overline S_1}^{|M_1|} = \Bigl(
\overline{\Lambda_0^2 m_1}\Bigr)^{|M_1|}
\Bigl({a_{12}\over \Lambda_0}\Bigr)^{|M_2|}
\Bigl({a_{31}\over \Lambda_0}\Bigr)^{-|M_3|}\, 
\exp(-2\pi i \alpha_1)
\cr
&{S_2}^{|M_2|} = 
\Bigl({\Lambda_0^2 m_2}\Bigr)^{|M_2|}
\Bigl({\overline{ a_{12}\over \Lambda_0}}
\Bigr)^{|M_1|}\Bigl({a_{23}\over \Lambda_0}\Bigr)^{-|M_3|} \,
\exp(-2\pi i \alpha_2)
\cr
&{S_3}^{|M_3|} = \Bigl({\Lambda_0^2 m_3}\Bigr)^{|M_3|}
\Bigl({a_{23}\over \Lambda_0}\Bigr)^{-|M_2|}
\Bigl(\overline{a_{31}\over \Lambda_0}\Bigr)^{-|M_1|}\,
\exp(-2\pi i \alpha_3)}.
$$
The $S_i$ are identified with the gaugino condensates of the low energy, $U(M_1)\times U(M_2)\times U(M_3)$ gauge theory.  The gaugino condensates are the order parameters of the low energy physics and as such should not depend on the cutoff $\Lambda_0$.  Let's then introduce three new confinement scales, $\Lambda_i$, defined as
$$
S_i = \Lambda_i^3.
$$
In fact, only two of these are independent.  As a consequence of homology relation \homtwo , the gauge couplings satisfy $\alpha_1+\alpha_2=\alpha_3$, which implies that
$$
\Bigl(\overline{{\Lambda_1\over
\Delta}}\Bigr)^{3|M_1|}
\Bigl({ \Lambda_2\over \Delta}\Bigr)^{3|M_2|}
=\Bigl({\Lambda_3\over \Delta}\Bigr)^{3|M_3|},
$$
where $\Delta$ is given in \DD .  Requiring that the scales $\Lambda_i$ do not depend on the cutoff scale, we can read off how the gauge couplings run with $\Lambda_0$,
\eqn\run{\eqalign{
g_{1}^{-2}(\Lambda_0) = &-
\log \Bigl(\overline{\Lambda_1^3\over
\Lambda_0^2 m_1
}\Bigr)^{|M_1|}-\log
\Bigl({ \Lambda_0\over a_{12}}
\Bigr)^{|M_2|}-\log\Bigl({\Lambda_0\over a_{13}}\Bigr)^{-|M_3|}\cr
g_{2}^{-2}(\Lambda_0)
 = &
-\log\Bigl({\Lambda_2^3\over
\Lambda_0^2 m_2}\Bigr)^{|M_2|}-\log
\Bigl({\overline{ \Lambda_0\over a_{12}}}
\Bigr)^{|M_1|}-\log\Bigl({\Lambda_0\over a_{23}}\Bigr)^{-|M_3|}.
}}
As was noticed in \Aganagic , this kind of running of the gauge couplings and relation between strong coupling scales is very similar to what occurs in the {\it supersymmetric} gauge theory (as studied in \vud ) obtained by wrapping $M_i$ branes of the same kind on the three $S^2$'s. The only difference is that branes and anti-branes lead to complex conjugate runnings, as if the spectrum of the theory remained the same, apart from the chirality of the fermions on the brane and the anti-brane getting flipped. This is natural, as the branes and the anti-branes have opposite GSO projections, so indeed a different chirality fermion is kept.  In addition, the open string RR sectors with one boundary on branes and the other on anti-branes has opposite chirality kept as well, and this is reflected in the above formulas.

To this order, the value of the potential at the critical point is
$$
V_* =\sum_{I} {|M_I|\over g_{I}^2} -\;
{1\over 2 \pi}|M_1||M_2|
\log\Bigl(
|{a_{12}\over \Lambda_0}|\Bigr)\;+\;{1\over 2 \pi}|M_1||M_3|
\log\Bigl(|{a_{13}\over \Lambda_0}|\Bigr)
$$
The first terms are just due to the tensions of the branes. The
remaining terms are due to the Coulomb and gravitational interactions
of the branes, which come from the one-loop interaction in the open
string theory.  There is no force between the $M_2$ branes wrapping
$S^2_2$ and the $M_3$ branes on $S^2_3$, since $M_{2,3}$ are both
positive, so the open strings stretching between them should be
supersymmetric.  On the other hand, the $M_1$ anti-branes on $S^2_1$
should interact with the $M_{2,3}$ branes as the Coulomb and gravitational
interactions should no longer cancel. This is exactly what one sees
above. The $M_1$ anti-branes on $S^2_1$ attract the $M_3$ branes on
$S^2_3$, while they repel the branes on $S^2_2$. We will see in the
next section that more generally, branes and anti-branes wrapping
2-cycles with negative intersection numbers (in the ALE space)
attract, and those wrapping 2-cycles with positive intersection
numbers repel.
Since\foot{The second relation is due to the self intersection numbers of $S^2_1$ and $S^2_2$ being $-2$.} 
$$
e_1\cdot e_2=1, \qquad e_1\cdot e_3 =-1, 
$$
this is exactly what we see here.

\subsec{$M_{1,2}>0$, $M_{3}<0$}

With only the non-simple root wrapped by anti-branes, the critical points of the potential now correspond to
$$
{S_1}^{|M_1|} = 
\Bigl({\Lambda_0^2 m_1}\Bigr)^{|M_1|}\Bigl({a_{12}\over \Lambda_0}
\Bigr)^{|M_2|}\Bigl({\overline{a_{31}\over \Lambda_0}}\Bigr)^{-|M_3|}\, 
\exp(-2\pi i \alpha_1)
$$
$$
{S_2}^{|M_2|} = 
\Bigl({\Lambda_0^2 m_2}\Bigr)^{|M_2|}
\Bigl({a_{12}\over \Lambda_0}
\Bigr)^{|M_1|}\Bigl({\overline{a_{23}\over \Lambda_0}}\Bigr)^{-|M_3|}\, 
\exp(-2\pi i \alpha_2)
$$
$$
{\overline{S_3}}^{|M_3|} = 
\Bigl({\overline{\Lambda_0^2 m_3}}\Bigr)^{|M_3|}
\Bigl({a_{23}\over \Lambda_0}\Bigr)^{-|M_2|}
\Bigl({a_{31}\over \Lambda_0}\Bigr)^{-|M_1|}\,
\exp(-2\pi i \alpha_3)
$$
In this case, the K\"ahler parameters $\alpha_{1,2}$ run as
$$\eqalign{
g_1^{-2}(\Lambda_0) =& -\log\Bigl(\overline{\Lambda_1^3\over
\Lambda_0^2 m_1}\Bigr)^{|M_1|}
-\log\Bigl({ \Lambda_0\over a_{12}}
\Bigr)^{|M_2|}-\log\Bigl({\Lambda_0\over a_{13}}\Bigr)^{-|M_3|}\cr
g_2^{-2}(\Lambda_0)
=& -\log
\Bigl({\Lambda_2^3\over
\Lambda_0^2 m_2}\Bigr)^{|M_2|}
-\log\Bigl({\overline{ \Lambda_0\over a_{12}}}
\Bigr)^{|M_1|}-\log\Bigl({\Lambda_0\over a_{23}}\Bigr)^{-|M_3|}}
$$ 
where
$$
\Bigl({\Lambda_1\over \Delta }\Bigr)^{3|M_1|}
\Bigl({ \Lambda_2\over \Delta}\Bigr)^{3|M_2|}
=\Bigl({\overline{\Lambda_3\over \Delta}}\Bigr)^{3|M_3|}.
$$
This follows the same pattern as seen in \Aganagic\ and in the previous subsection.  The branes and anti-branes give complex conjugate runnings, as 
do the strings stretching between them.
 
The value of potential at the critical point is, to this order,
$$
V_* =\sum_{I} {|M_I|\over g_{I}^2} +\;
{1\over 2 \pi}|M_1||M_3|\,
\log\Bigl(|{a_{13}\over \Lambda_0}|\Bigr)\;+\;{1\over 2 \pi}|M_2||M_3|\,
\log\Bigl(|{a_{23}\over \Lambda_0}|\Bigr).
$$
Again, the first terms are universal, coming from the brane tensions.  The remaining terms are the one-loop interaction terms. There is no force between the $M_1$ branes wrapping $S^2_1$ and the $M_2$ branes on $S^2_2$, since now both $M_{1,2}$ have the same sign.
The $M_3$ anti-branes on $S^2_3$ attract both $M_1$ branes on $S^2_1$ and the $M_2$ branes on $S^2_2$, since, in the ALE space
$$
e_1\cdot e_3 =e_2\cdot e_3=-1.
$$

In the next subsection, we will show that both of these brane/anti-brane systems are perturbatively stable for large separations.

\subsec{Metastability}

The system of branes and anti-branes engineered above should be
perturbatively stable when the branes are weakly interacting -- in
particular, at weak 't Hooft coupling. The open/closed string duality
implies that the dual closed string vacuum should be metastable as
well. In this subsection, we'll show that this indeed is the case.
Moreover, following \Seo , we'll show that perturbative stability is
lost as we increase the 't Hooft coupling.  While some details of this section will be specific to the $A_2$ case discussed above, the general aspects of the analysis will be valid for any of the ADE fibrations discussed in the next section.

To begin with, we note that the equations of motion, derived from the potential \pot , are
\eqn\twoeom{
\partial_k V = {-1 \over 2i} {\cal F}_{kef}G^{ae}G^{bf}(\alpha_a
+ M^c \bar\tau_{ac})(\bar\alpha_b + M^d \bar\tau_{bd}) = 0,
}
and moreover, the elements of the Hessian are 
\eqn\secondderiv{
\eqalign{
\del_p\del_q V=& G^{ia} G^{bj}i {\cal F}_{abpq}(\alpha_i+M^k\bar\tau_{ki})(\bar\alpha_j+M^r\bar\tau_{rj})\cr
&+2 G^{ia}G^{bc}G^{dj}i {\cal F}_{abp}i {\cal F}_{cdq}(\alpha_i+M^k\bar\tau_{ki})(\bar\alpha_j+M^r\bar\tau_{rj})\cr
\del_q\del_{\bar q}V
=&-G^{ia}G^{bc}G^{dj}i {\cal F}_{abp}i {\bar {\cal F}}_{cdq}(\alpha_i+M^k\bar\tau_{ki})(\bar\alpha_j+M^r\tau_{rj})\cr
&-G^{ic}G^{da}G^{bj}i{\cal F}_{abp}i{\bar {\cal F}}_{cdq}(\alpha_i+M^k\tau_{ki})(\bar\alpha_j+M^r\bar\tau_{rj})
}
}
where we have denoted $\partial_c \tau_{ab} = {\cal F}_{abc}$, and similarly for higher derivatives of $\tau$.

In the limit where all the 't Hooft couplings $g_i^2 N_i$ are very small, the sizes of the dual 3-cycles 
$
S_a = \Lambda_a^3
$ 
are small compared to the separations between them, so we can keep
only the leading terms in the expansion of ${\cal F}_0$ in powers of
$S$, i.e., the one-loop terms in the matrix model. At one-loop, the
third and fourth derivatives of the prepotential are nonzero only if
all of the derivatives are with respect to the same variable.
Expanding about the physical solution to this order,
\eqn\so{
\alpha_{a}+\sum_{M_b>0}\tau_{ab} M_b + 
\sum_{M_b<0}{\overline \tau}_{ab} M_b =0.
}
The non-vanishing elements of the Hessian are
\eqn\hess{
\eqalign{
\del_i\del_j V={1\over4\pi^2}{|M^iM^j|\over S_iS_j}G^{ij}\qquad &
i,j\;\;opposite\;type\cr
\del_i\del_{\bar j} V={1\over4\pi^2}{|M^iM^j|\over S_i\bar S_j}G^{ij}\qquad &i,j\;\;same\;type
}
}
where the `type' of an index refers to whether it corresponds to branes or anti-branes.

To get a measure of supersymmetry breaking, consider the fermion 
bilinear couplings. Before turning on fluxes, the theory has ${\cal N}=2$ supersymmetry, and the choice of superpotential \supp\ breaks this explicitly to ${\cal N}=1$. For each 3-cycle, we get a 
chiral multiplet $(S_i,\psi_i)$ and a vector multiplet 
$(A_i, \lambda_i)$ where $\psi_i, \lambda_i$ are a pair of Weyl 
fermions. It is easy to work out \Aganagic\ that the 
coefficients of the non-vanishing fermion bilinears 
are
$$\eqalign{
m_{\psi^a\psi^b} &= 
{1\over 2} G^{cd}(\alpha_d + M^{e}{\bar \tau}_{de}){\cal F}_{abc},\cr
m_{\lambda^a\lambda^b}& = {1\over 2} G^{cd}({\bar \alpha}_d + M^{e}{\bar \tau}_{de}){\cal F}_{abc},
}
$$
and evaluating this in the vacuum we find
$$\eqalign{
m_{\psi^a\psi^b} &=  -{1\over 4 \pi}{M_a + |M_a| \over S_a}{\delta}_{ab},  \cr
m_{\lambda^a\lambda^b} & =  -{1\over 4\pi} 
{M_{a}-|M_a| \over S_a} \delta_{ab}.}
$$
Bose-Fermi degeneracy is restored in the limit where we take
$$
(G_{ij})^2/G_{ii} G_{jj}\ll 1,\qquad i,j\;\;opposite\;type.
$$
In this limit we get a decoupled system of branes and anti-branes
except that for nodes wrapped with branes, $S_a$'s get paired up with
$\psi's$, and for nodes wrapped with anti-branes they pair with $\lambda$'s, corresponding to a different half of ${\cal N}=2$ supersymmetry being preserved in the two cases.\foot{The kinetic terms of both bosons and fermions are computed with the same metric $G_{ab}$.} This is the limit of extremely weak 't Hooft coupling, and the sizes of the cuts are the smallest scale in the problem by far
\eqn\susylimit{
{\Lambda_i\over \Delta}\ll {a_{ij}\over \Lambda_0},\; {\Delta\over\Lambda_0}<1.
}
In this limit the Hessian is manifestly positive definite.  
In fact the Hessian is positive definite as long as the one-loop approximation is valid. To see this note that the
determinant of the Hessian is, up to
a constant, given by
\eqn\det{
{\rm Det}( \del^2V) \sim\left({1\over {\rm Det}\,G}\prod_{i=1}^n\left|{M^i\over S_i}\right|^2\right)^2.
}
It is never zero while the metric remains positive definite, so a
negative eigenvalue can never appear.  Thus, one can conclude that as
long as all the moduli are in the regime where the 't Hooft couplings
are small enough for the one-loop approximation to be valid, the
system will remain stable to small perturbations.

Let's now find how the solutions are affected by the inclusion of
higher order corrections. At two loops, an exact analysis of
stability becomes difficult in practice. However, in various limits
one can recover systems which can be understood quite well. For
simplicity, we will assume that the $\alpha_i$ are all pure imaginary,
and all the parameters $a_{ij}$ and $\Lambda_0$ are purely real. Then
there are solutions where the $S_i$ are real. In appendix B, we
show that in this case, upon including the two-loop terms, the
determinant of the Hessian becomes
\eqn\twodetmm{
\left( {\rm Det} \  G^{ab}\right)^2 \left(\prod_c {|M^c| \over i {\cal
F}_{ccc}}\right)^4 {\rm Det}\left(\delta_{cb} + 
 G_{cb} {i{\cal F}_{bbbb} \delta^b \over  
i{\cal F}_{bbb}i{\cal F}_{ccc}  |M^c|} \right)
{\rm Det}\left(\delta_{cb} - 
 G_{cb} {i{\cal F}_{bbbb} \delta^b \over  
i{\cal F}_{bbb}i{\cal F}_{ccc}  |M^c|} \right)
}
where
\eqn\twodeltsolsol{
\delta^k = {1 \over 2 |M^k|{\cal F}_{kkk}} {\cal F}_{kab}(- |M^a| |M^b| + M^a M^b)
}
and $\delta_{cb}$ is the Kronecker delta. The first two terms in
\twodetmm\ never vanish, since the metric has to remain positive
definite, so we need only analyze the last two determinants. We can
plug in the one-loop values for the various derivatives of the
prepotential, and in doing so obtain
\eqn\twodetdet{
{\rm Det}\left( \delta_{ab} \pm 2\pi G_{ab} {S^a \over \Delta^3} {x^b
\over |M^a|} \right) = 0
} 
with either choice of sign. Above, we have rewritten
eq. \twodeltsolsol\ as
\eqn\twodeltrev{
\delta^a = {S^a \over \Delta^3 } x^a.
}
This is a convenient rewriting because $S/\Delta^3$ is the parameter
controlling the loop expansion, and $x^a$ is simply a number which
depends on the $N^i$ but no other parameters.

Consider the case where, for some $i$, 
a given $S^3_i$ grows much larger than the other two.
We can think of this as increasing the {\it effective} 't Hooft coupling for that node, or more precisely, increasing
$$
\Bigr({\Lambda_{M_i}\over\Delta}\Bigl)^3 = \exp{\Bigl(-{1\over| M_i| 
g^{2}_{i,eff}(\Delta)}\Bigr)}.
$$
Recall that the two-loop equations of
motion for $real$ $S_i$, are given by
\eqn\neoms{
g_{i, eff}^{-2}(\Delta)=
-|M_i|\log (|{S_i\over\Delta^3}|)+G_{ik}\delta^k
}
where
$${\eqalign{
g_{1, eff}^{-2}(\Delta)&= g_{1}^{-2}(\Lambda_0) - |M_1|(L_{12}+L_{13})+|M_2|L_{12}-|M_3|L_{13}\cr
g_{2, eff}^{-2}(\Delta)&=g_{2}^{-2}(\Lambda_0) - |M_2|(L_{12}+L_{23})+|M_1|L_{12}-|M_3|L_{13}\cr
g_{3, eff}^{-2}(\Delta)&=
g_3^{-2}(\Lambda_0) - |M_3|(L_{13}+L_{23})-|M_1|L_{13}-|M_2|L_{23}.}}
$$
Here we've adopted the notation
$L_{ij}=\log{\Lambda_0\over a_{ij}}$ and the $\delta^k$ are as defined
in \twodeltsolsol.  
Note that in each case, two of the equations can be solved straight
off.  It is the remaining equations which provide interesting
behavior and can result in a loss of stability.
Correspondingly, the vanishing of the Hessian determinant in \twodetmm\ is then equivalent to the vanishing of
its $ii$ entry (where we have assumed a vacuum at real $S$):
\eqn\ssimoo{
1 \pm G_{ii}\,{S_i \over \Delta^3}{x_i \over |M_i|} =0.
}
We'll see that we can approximate
$$
G_{ii} = -\log(|{S_i\over \Delta^3}|)+ L_{i}\sim 
L_i
$$
where we have defined 
$$
L_i = L_{ij}+L_{ik}, \qquad
i\neq j\neq k,
$$
so this provides the following conditions:
\eqn\conditions{
\pm 1=L_i {x_i\over|M_i|}{S_i\over\Delta^3}.
}
The above equation, taken with positive sign, is equivalent to
the condition for stability being lost by setting the
determinant of the gradient matrix of the equations to zero.  
The equation with minus sign comes from losing stability in imaginary direction. Correspondingly, the equation of motion for the one
node with growing 't Hooft coupling becomes
\eqn\nodeone{
g_{i, eff}^{-2}=-|M_i|\log {S_i\over\Delta^3}+ L_i x_i {S_i\over\Delta^3}. 
}
One of the equations \conditions\ must be solved in 
conjunction with \nodeone\ if stability is to be lost.  

The sign of $x_i$ can vary depending on the specifics of the
charges. 
In all the cases, as the effective 't Hooft coupling increases, solutions move to larger values of $S_{i}$.  For sufficiently large
values, in the absence of some special tuning of the charges,
\conditions\ will be satisfied for one of the two signs.  The only
question then is whether the $S_{i}$ can get large enough, or whether
a critical value above which the equation of motion can no longer be
solved is reached {\it before} an instability sets in. In the equation above,  if $x_i$ is negative, then there will be no such
critical value, and $S_i$ can continue to grow unbounded.
Correspondingly, a large enough value of the 't Hooft coupling can always be reached where \conditions\ is satisfied with negative sign.
Alternatively, if the coefficient $x_i$ is positive, there will be a
critical value for $L_i$ at which the right hand side of the equation
takes a minimum value. This occurs at 
$ 
({S_{i,*}/ \Delta^3})={|M_i|\over x_i L_i}, 
$
which is precisely \conditions\ with positive left hand side.
So, for $any$ value of $x_i$ an instability develops at finite
effective 't Hooft coupling corresponding to
$$ 
{S_{i,*}\over \Delta^3}={|M_i|\over |x_i| L_i}, 
$$
or more precisely, at 
$$
|M_i| g_{i,eff}^{2}(\Delta) =\log^{-1}({|M_i|\over x_i L_i}).
$$ 
This critical value of the effective 't Hooft coupling
can be achieved by increasing the number of branes on that node, or, in case of nodes one and two, by letting the corresponding {\it bare} 't Hooft coupling increase. This is true as long as supersymmetry is broken and the corresponding two-loop correction is non-vanishing, i.e. as long as $x_i\neq 0$. It is reasonable to suspect that in the degenerate case, where charges conspire to set $x_i$ to zero even with broken supersymmetry, 
the instability would set in at three loops.

It is natural to ask the fate of the system once metastability is
lost. It should be the case \Seo\ that it rolls to another a critical point corresponding to
shrinking the one compact $B$-type cycle, $\CB_1+\CB_2-\CB_3$. 
To describe this point in the moduli space, introduce a new basis of periods in which this shrinking $B$-cycle becomes one of the $A$ periods:
$$
\eqalign{\oint_{\CA_1'} H =M_1+M_3,\qquad \oint_{\CA_2'}H&=M_2+M_3 ,\qquad
\oint_{\CA_3'}H=0,\cr
\int_{\CB_1'}H= \alpha_1,\qquad \int_{\CB_2'} H &= \alpha_2,\qquad \int_{\CB_3'}H=M_3,
}
$$
where $H=H^{RR} + \tau H^{NS}$. In particular, there is no flux through the new cycle $\CA_3'$. In fact, by setting $M_1=M_2=-M_3
=M$, there is no flux through any of the $\CA'$ cycles.\foot{In the more general case, the system should be attracted to a point where only $\CA_3'$ shrinks.} 
For $S'_i=\int_{\CA'}\Omega$ sufficiently large that we can ignore the light D3 branes wrapping this cycle,
$$ \tau'_{ii} \sim {1\over 2\pi i}\log {S_i'\over\Delta^3}, \qquad \tau'_{i\neq j} \sim const,$$
it is easy to see that the system has an effective potential 
that would attract it to the point where the $S'_i=0$ and the cycles shrink:
$$V_{eff} \sim V_0+ \sum_i \Bigl|{c_i \over \log{|{S'_i\over\Delta^3}|}}\Bigr|^2
$$ 
where $c_i \sim \int_{\CB_i'}H$. 
By incorporating the light D3 branes wrapping the flux-less, shrinking 
cycles, the system would undergo a geometric
transition to a non-K\"ahler manifold \strominger .  There, the cycle shrinks and a
new 2-cycle opens up, corresponding to condensing a D3 brane
hypermultiplet.  However, this 2-cycle becomes the boundary of a compact 3-cycle $\CB'$ which get punctured in the transition that shrinks the $\CA'$ cycles. 
A manifold where such a 2-cycle has nonzero volume is automatically non-K\"ahler, but it is supersymmetric.  As we'll review shortly, the shrinking cycle $\CA_3'$ is also the cycle wrapped by the D5 brane domain walls that mediate the non-perturbative decay of the metastable
flux vacua. The loss of metastability seems to be correlated with
existence of of a point in the moduli space where the domain walls
become light and presumably fluxes can annihilate classically (this
also happened in the $A_1$ model studied in \Seo ).
In particular, in the last section of this paper, we'll provide two
examples of a system where the corresponding points in the complex
structure moduli space are {\it absent}, but which are exactly stable
perturbatively even though they are non-supersymmetric (one of them will be stable non-perturbatively as well).
It must be added, as discussed in \Seo , that it is far from clear whether the light domain walls can be ignored,
and so whether the system truly rolls down to a supersymmetric vacuum. 
A more detailed analysis of the physics at this critical point is beyond the scope of this paper.

It was suggested in \Seo\ that the loss of stability might be
related to the difference in the value of $V_*$ between the starting
vacuum and a vacuum to which it might tunnel becoming small, and thus 
the point where Coulomb attraction starts to dominate in a subset of branes.  
In the more complicated geometries at hand, it seems that such a simple statement does not carry over.
This can be seen by noting that, for certain configurations of brane
charges in our case, an instability can be induced without having any
effect on the $\Delta V_*$ between vacua connected by tunneling
events. We are led to conclude that the loss of stability is a strong
coupling effect in the non-supersymmetric system, which has no simple
explanation in terms of our open string intuition. This should have perhaps been clear, in that the point to which the system apparently rolls has no straightforward explanation in terms of brane annihilation.
 
\subsec{Decay Rates}

We now study the decays of the brane/anti-brane systems of the
previous section. This closely parallels the analysis of \Aganagic .
We have shown that when the branes and anti-branes are sufficiently
well-separated, the system is perturbatively stable.
Non-perturbatively, the system can tunnel to lower energy vacua, if
they are available.  In this case, the available vacua are constrained
by charge conservation -- any two vacua with the same net charges
$$
N_1 = M_1+M_3, \qquad 
N_2 = M_2+M_3
$$
are connected by finite energy barriers.  The false vacuum decay proceeds by the nucleation of a bubble of lower energy vacuum. 

The decay process is easy to understand in the closed string language.  The vacua are labeled by the fluxes through the three $S^3$'s
$$
\int_{\CA_I} H^{RR} = M_I. \qquad I=1,2,3
$$
Since RR 3-form fluxes jump in going from the false vacuum to the true vacuum, the domain walls that interpolate between the vacua are D5 branes.  Over a D5 brane wrapping a compact 3-cycle $C$ in the Calabi-Yau, the fluxes jump by an amount
$$
\Delta M_I =  \# (C\cap \CA_I)
$$
In the present case, it is easy to see that there is only one compact 3-cycle $C$ that intersects the $A$-cycles,
$$
C = \CB_1+\CB_2-\CB_3.
$$
So, across a D5 brane wrapping $C$, the fluxes through $\CA_{1,2}$ decrease by one unit, and the flux through $\CA_3$ increases by one unit. Note that this is consistent with charge conservation for the branes. In fact, the domain walls in the open and the closed picture are essentially the same.  In the open string language, the domain wall is also a D5 brane, but in this case it wraps a three-chain obtained by pushing $C$ through the geometric transition.  The three-chain has boundaries on the minimal $S^2$'s, and facilitates the homology relation \homtwo\ between the 2-cycles.  

The decay rate $\Gamma$ is given in terms of the action $S_{inst}$ of the relevant instanton.
$$
\Gamma \sim \exp(-S_{inst})
$$
Since the Calabi-Yau we have been considering is non-compact, we can neglect gravity, and the instanton action is given by
$$
S_{inst} = {27 \pi \over 2} {S_D^4 \over (\Delta V_*)^3}
$$
where $S_D$ is the tension of the domain wall, and ${\Delta V_*}$ is the change in the vacuum energy across the domain wall.  While this formula was derived in \coleman\ in a scalar field theory, it is governed by energetics, and does not depend on the details of the theory as long as the semi-classical approximation is applicable.  

In the present case, the tension of the domain wall is bounded below by
\eqn\dom{
S_D = {1\over g_s} 
\int_{C} \Omega,
}
since the $\int_{C} \Omega$ computes the lower bound on the volume of any 3-cycle in this class, and the classical geometry is valid to the leading order in $1/N$, the order to which we are working.  The tension of the domain wall is thus the same as the tension of a domain wall interpolating between the supersymmetric vacua, and to leading order (open-string tree-level) this is given by the difference between the tree-level superpotentials \supp 
$$
\int_{C} \Omega \sim W_3(a_3) - W_1(a_1)-W_2(a_2)= {1\over 2} \Delta^3,
$$
where $\Delta^3$ is defined in \DD . This is just the ``holomorphic area'' of the triangle in figure 1.  The area is large as long as all the brane separations are large, and as long as this is so, it is independent of the fluxes on the two sides of the domain wall.  

At the same time, the difference in the potential energy between the initial and the final states is given by the classical brane tensions,
$$
\Delta V = V_i-V_f = \sum_{I} (|M_{I}|-|M_{I}'|)/g_{I}^2. 
$$
The fate of the vacuum depends on the net charges. If $N_{1,2}$ are both positive, then the true vacuum is supersymmetric. Moreover, there is a landscape of degenerate such vacua, corresponding to all possible ways of distributing branes consistent with charge conservation such that $M_I'$ are all positive.  Starting with, say, $(M_1,M_2,M_3) = (N_1+k,N_2+k,-k)$, where $k>0$, this can decay to $(N_1,N_2,0)$ since 
$$
\Delta V = V_i-V_f= 2 {k|r_3|\over g_s}, 
$$
corresponding to $k$ branes on $S^2_3$ getting annihilated, where $r_3$ is the K\"ahler area of $S^2_3$.  The decay is highly suppressed as long as string coupling $g_s$ is weak and the separation between the branes is large. The action of the domain wall is $k$ times that of \dom\foot{All quantities being measured in string units.}, so
\eqn\nin{
S_{inst} = {27 \pi \over 32} {k\over g_s}{|\Delta|^{12}\over |r_3|^3}={27\pi\over32}{|g_3|^3\over g_s^4}k|\Delta|^{12}.
}
The instanton action \nin\ depends on the cutoff scale $\Lambda_0$ due
to the running of the gauge coupling $g_3^{-2}(\Lambda_0)$. The
dependence on $\Lambda_0$ implies \MarsanoFE\ that \nin\ should be
interpreted as the rate of decay corresponding to fluxes decaying in
the portion of the Calabi-Yau bounded by $\Lambda_0$. 

If instead we take say $N_1>0>N_2$, then the lowest energy state corresponds to $N_1$
branes on node 1, $N_2$ anti-branes on node 2, with node 3 unoccupied. This is the case
at least for those values of parameters corresponding to the system being weakly coupled.
In this regime, this particular configuration gives an example of an exactly stable, non-supersymmetric vacuum in string theory -- there is no other vacuum with the same charges that has lower energy.
Moreover, as we'll discuss in section 6, for some special values of the parameters $m_{1,2}$ the system is exactly solvable, and can be shown to be $exactly$ stable even when the branes and the anti-branes are close to each other.

\newsec{Generalizations} 
Consider now other ADE fibrations over the complex plane.  As in \fibr\ we start with 
the deformations of 2-complex dimensional ALE singularities:

$${\eqalign{ A_k\;:\qquad& x^2+y^2+z^{k+1}=0\cr
D_r\; :\qquad & x^2 + y^2z
+ z^{r-1} = 0\cr
E_6\;: \qquad & x^2+y^3+z^4 =0\cr 
E_7\;: \qquad &
x^2+y^3+yz^3 =0\cr
E_8\;: \qquad & x^2+y^3+z^5 =0
}} 
$$ 
and fiber these over the complex $t$ plane, allowing the coefficients parameterizing the deformations to be $t$ dependent.  The requisite deformations of the singularities are canonical (see \ckv\ and references therein). For example, the deformation of the $D_r$ singularity is 
$$
x^2+y^2z+z^{-1}(\prod_{i=1}^r (z-z_i^2)-\prod_{i=1}^r z_i^2)+2\prod_{i=1}^rz_i\,y.
$$
In fibering this over the $t$ plane, the $z_i$ become polynomials $z_i(t)$ in $t$.\foot{This is the so called ``non-monodromic'' fibration.  The case where the $z_i$ are instead multi-valued functions of $t$ corresponds to the ``monodromic'' fibration \ckv .}  After deformation, at a generic point in the $t$ plane, the ALE space is smooth, with singularities resolved by a set of $r$ independent 2-cycle classes
$$
S^2_i, \qquad i=1,\ldots r
$$
where $r$ is the rank of the corresponding Lie algebra. The 2-cycle classes intersect according to the ADE Dynkin diagram of the singularity:
\bigskip
\centerline{\epsfxsize 3.0truein\epsfbox{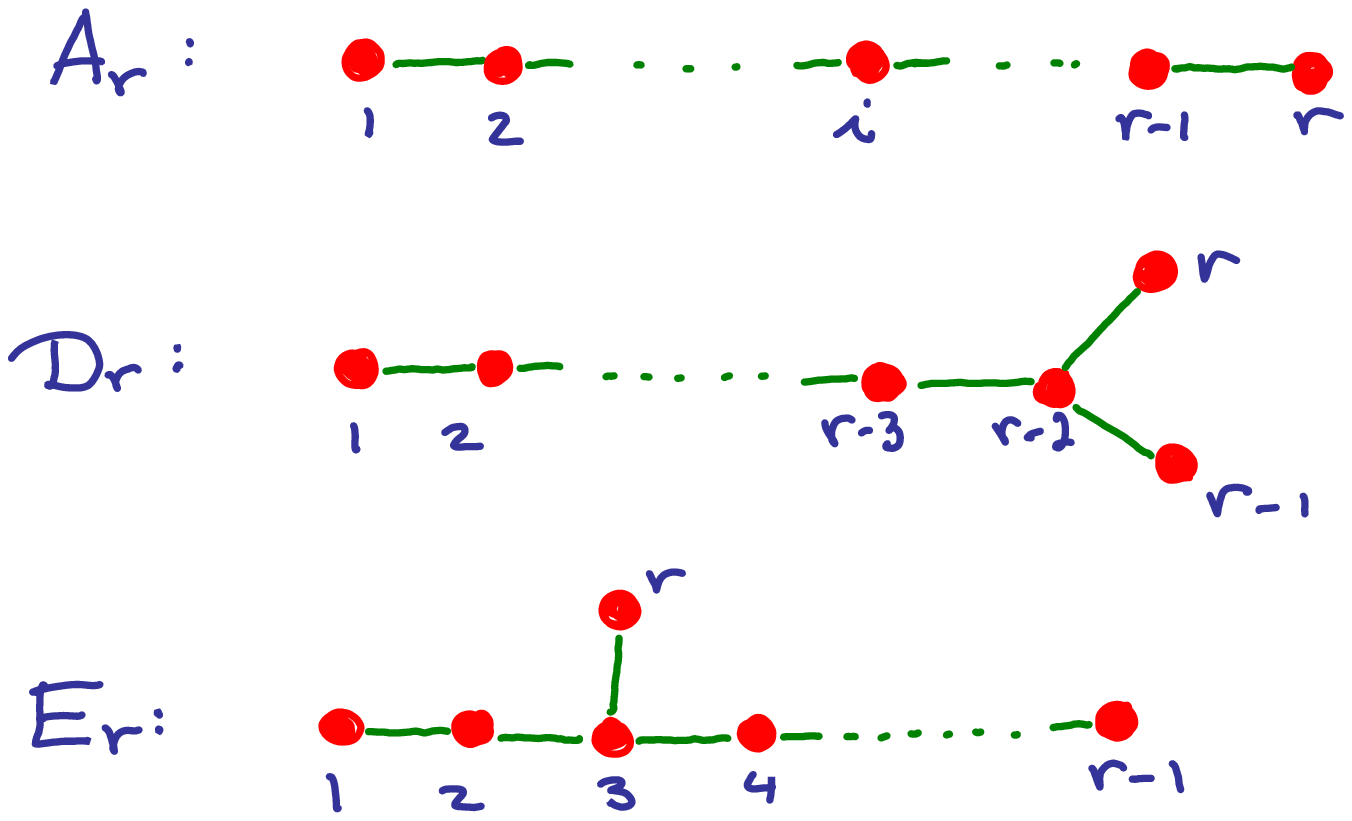}}

\noindent{\ninepoint
\baselineskip=2pt {\bf Fig. 2.} {Dynkin diagrams of the ADE Lie algebras.  Every node corresponds to a simple root and to a 2-cycle class of self intersection $-2$ in the ALE space.  The nodes that are linked correspond to 2-cycles which intersect with intersection number 
$+1$.}}
\bigskip
The deformations can be characterized by ``superpotentials'', 
$$
W'_{i}(t) = \int_{S_{i,t}^2} \omega^{2,0}.
$$
which compute the holomorphic volumes of the 2-cycles at fixed $t$.  For each positive root $e_{I}$, which can be expanded in terms of simple roots $e_i$ as 
$$
e_{I} = 
\sum_{I} n_{I}^i e_i
$$ 
for some positive integers $n_I^i$, one gets a zero-sized, primitive 2-cycle at points in the $t$-plane where
\eqn\pos{
W'_I(t) =\sum_{i} n_{I}^i\,W'_i(t)=0.
}
Blowing up the singularities supplies a minimal area to the 2-cycles at solutions of \pos ,
$$
t=a_{I,p},
$$
where $I$ labels the positive root and $p$ runs over all the solutions to \pos\ for that root.  

As shown in \ckv\ and references therein, the normal bundles to the minimal, holomorphic $S^2$'s obtained in this way are always ${\cal O}(-1)\oplus {\cal O}(-1)$, and correspondingly the $S^2$'s are isolated.\foot{In \ckv\ the authors also considered the monodromic ADE fibrations, where the 2-cycles of the ALE space undergo monodromies around paths in the $t$ plane. In this case, the novelty is that the $S^2$'s can appear with normal bundles ${\cal O}\oplus {\cal O}(-2)$ or ${\cal O}(-1)\oplus {\cal O}(3)$. Wrapping branes and anti-branes on these cycles is not going to give rise to new metastable vacua, since there will be massless deformations moving the branes off of the $S^2$'s. It would be interesting to check this explicitly in the large $N$ dual.} 
This implies that when branes or anti-branes are wrapped on the $S^2$'s, there is an energy cost to moving them off. Moreover, the parameters that enter into defining the $W_i$, as well as the K\"ahler classes of the $S^2$'s, are all non-dynamical in the Calabi-Yau.  As a consequence, if we wrap branes and anti-branes on minimal $S^2$'s, the non-supersymmetric system obtained is metastable, at least in the regime of parameters where the $S^2$'s are well separated.

The ALE fibrations have geometric transitions in which each minimal $S^2$ is replaced by a minimal $S^3$. A key point here is that none of the 2-cycles have compact, dual 4-cycles, so the transitions are all locally conifold transitions.  The one-loop prepotential ${\cal F}_0$ for all these singularities was computed in \vud , and is given by
\eqn\ADE{\eqalign{
2\pi i {\cal F}_0(S) = &\half\sum_{b} \;{S_{b}^2} \,
\Bigl(\log\Bigl({S_{b}\over W''_{I}(a_{b})\,\Lambda_0^2}\Bigr)-{3\over 2}\Bigr)
+
\half\sum_{b\neq c}\,{{e_{I(b)}\cdot e_{J(c)}}}\,
S_{b}\,S_{c}\,
\log\Bigl({a_{bc}\over \Lambda_0}\Bigr),
}}
where the sum is over all critical points $b=(I,p)$, and $I(b)=I$ denotes the root $I$ to which the critical point labeled by $b$ corresponds.  We are neglecting cubic and higher order terms in the $S_{I,p}$, which are related to higher loop corrections in the open string theory.  Above, $W_I(t)$ is the superpotential corresponding to the root $e_I$, and $e_{I}\cdot e_{J}$ is the inner product of two positive, though not necessarily simple, roots.  Geometrically, the inner product is the same as $minus$ the intersection number of the corresponding 2-cycles classes in the ALE space.

Consider wrapping $M_{b}$ branes or anti-branes on the minimal $S^{2}$'s labeled by $b=(I,p)$. We'll take all the roots to be positive, so we get branes or anti-branes depending on whether $M_{b}$ is positive or negative.  The effective superpotential for the dual, closed-string theory is given by \effsupp . From this and the corresponding effective potential \pot , we compute the expectation values for $S_b$ in the metastable vacuum to be
$$\eqalign{
&{\overline{S_{b}}}^{|M_{b}|} = 
\Bigl({\overline{\Lambda_0^2 \;{W}_{I}''(a_{b})}}\Bigr)^{|M_{b}|}
\prod^{M_{c}<0}_{b\neq c}
\Bigl(\overline{a_{bc}\over \Lambda_0}\Bigr)^{|M_{c}|}
\prod^{M_c>0}_{c}
\Bigl({a_{bc}\over \Lambda_0}\Bigr)^{|M_{c}|}
\, \exp(-2\pi i \alpha_{I(b)}), \qquad M_b<0
\cr
&{S_{b}}^{|M_{b}|} = 
\Bigl({\Lambda_0^2 \;{W}_{I}''(a_{b})}\Bigr)^{|M_{b}|}
\prod^{M_{c}>0}_{b\neq c}
\Bigl({a_{bc}\over \Lambda_0}\Bigr)^{|M_{c}|}
\prod^{M_c<0}_c
\Bigl(\overline{a_{bc}\over \Lambda_0}\Bigr)^{|M_{c}|}
\, \exp(-2\pi i \alpha_{I(b)}). \qquad M_b>0
}
$$
The value of the effective potential at the critical point is given by
$$
V_{*} = \sum_{b}{|M_{b}|\over g_{I(b)}^2} + 
\sum^{M_{b}>0>M_{c}}_{b,c}{1\over2\pi} {e_{I(b)}}\cdot{e_{J(c)}}\,
\log\Bigl(\Bigl|{a_{bc}\over \Lambda_0}\Bigr|\Bigr).
$$
The first term in the potential is just the contribution of the tensions of all the branes and anti-branes. The second term comes from the Coulomb and gravitational interactions between branes, which is a one-loop effect in the open string theory. As expected, at this order only the brane/anti-brane interactions affect the potential energy. The open strings stretching between a pair of (anti-)branes, are supersymmetric, and the (anti-)branes do not interact. The interactions between branes and anti-branes depend on
$$
e_I \cdot e_J 
$$
which is {\it minus} the intersection number -- in the ALE space -- of the 2-cycle classes wrapped by the branes. The branes and anti-branes attract if the 2-cycles they wrap have {\it negative} intersection, while they repel if the intersection number is {\it positive}, and do not interact at all if the 2-cycles do not intersect.  

For example, consider the $A_k$ quiver case, and a set of branes and anti-branes wrapping the 2-cycles obtained by blowing up the singularities at
$$
z_i(t) = z_j(t), \qquad z_{m}(t)=z_{n}(t)
$$
where $i<j$ and $m<n$.
The branes do not interact unless $i$ or $j$ coincide with either $m$ or $n$.  The branes attract if $i=m$ or $j=n$, in which case the intersection is either $-1$ or $-2$, depending on whether one or both of the above conditions are satisfied.  This is precisely the case when the branes and anti-branes can at least partially annihilate. If $j=m$ or $i=n$, then the 2-cycles have intersection $+1$, and the branes repel.  In this case, the presence of branes and anti-branes should break supersymmetry, but there is a topological obstruction to the branes annihilating, even partially. In fact, in the $A_k$ type ALE spaces, this result is known from the direct, open string computation \refs{\MukhiDN,\MukhiTE}. The fact that the direct computation agrees with the results presented here is a nice test of the conjecture of \Aganagic . 

\newsec{A Non-Supersymmetric Seiberg Duality}
In the supersymmetric case, with all $M_I$ positive, the engineered
quiver gauge theories have Seiberg-like dualities. In string theory,
as explained in \vud , the duality comes from an {\it intrinsic} ambiguity in how we resolve the ADE singularities to 
formulate the brane theory.\foot{The idea that Seiberg dualities have a geometric interpretation in string theory goes back a long while, see for example \refs{\Eli\Elit\Feng\Beasley-\Berenstein}. The fact that these dualities 
arise dynamically in string theory has for the first time been
manifested in \refs{\vud,\KS}.} The different resolutions are related by flops of the $S^2$'s
under which the charges of the branes, and hence the ranks of the
gauge groups, transform in non-trivial ways.  The RG flows, which are
manifest in the large $N$ dual description, force some of the $S^2$'s
to shrink and others to grow, making one description preferred over
the others at a given energy scale. In this section, we argue that
Seiberg dualities of this sort persist even when some of the
branes are changed to anti-branes and supersymmetry is broken.

\subsec{Flops as Seiberg Dualities}

For a fixed set of brane charges, one can associate different
Calabi-Yau geometries. There is not a unique way to blow up the
singularity where an $S^2$ shrinks, and the different blowups are
related by flops that shrink some 2-cycles and grow others. Instead of
giving a 2-cycle class $S^2_i$ a positive K\"ahler volume
$$
r_i = \int_{S^2_i} B^{NS}
$$
we can give it a $negative$ volume, instead. This can be thought of as replacing the 2-cycle class by one of the opposite orientation
$$
S^2_i\rightarrow {\tilde S}^2_i = -S^2_i.
$$
The flop of a simple root $S^2_i$ acts as on the other roots as a Weyl reflection which permutes the positive roots
\eqn\flopc{ S^2_j\rightarrow {\tilde S}^2_j = S^2_j - (e_j \cdot
e_i)\; S^2_i. 
}
The net brane charges change in the process, but in a way consistent with charge conservation
\eqn\larggen{
\sum_{i} N_i \, S^2_i = \sum_{i} {\tilde N}_i\,  {\tilde S}^2_i.
}
We can follow how the number of branes wrapping the minimal 2-cycles change in this process.  If $i$ is the simple root that gets flopped\foot{Flopping non-simple roots can be thought of in terms of a sequence of simple node flops, as this generates the full Weyl group.}, then $M_{i,p}$ goes to ${\tilde M}_{i,p}=-M_{i,p}$ and for other roots labeled by $J\neq i$
\eqn\smallgen{
M_{J,p} = {\tilde M}_{w(J),p}
}
where $w(J)$ is the image of $J$ under the Weyl group action.  

The size of the wrapped $S^2$ is proportional to the inverse gauge coupling for the theory on the wrapping branes,
\eqn\invg{
{g_i^{-2}(t)} \propto {1\over g_s} \int_{S^2_{i,t}} B_{NS}, 
}
so the flop \flopc\ transforms the gauge couplings according to
\eqn\flopgc{ g_j^{-2} \rightarrow {\tilde g}_j^{-2} = g^{-2}_j - (e_j \cdot
e_i)\; g^{-2}_i.
}
Generally, there is one preferred description for which the gauge
couplings are all positive.  In the geometry, we have the freedom to
choose the sizes of the 2-cycles $S^2_{i,t}$ at some fixed high
scale, but the rest of their profile is determined by the one-loop
running of the couplings \run\ throughout the geometry and by the
brane charges. The most invariant way of doing this is to specify the
scales ${\Lambda}_{i}$ at which the couplings \invg\ become strong.
We can then follow, using holography, the way the $B$-fields vary over
the geometry as one goes from near where the $S^3$'s are minimal,
which corresponds to low energies in the brane theory, to longer
distances, far from where the branes were located, which corresponds
to going to higher energies.  The $S^2$'s have finite size and shrink
or grow depending on whether the gauge coupling is increasing or
decreasing. We'll see that as we vary the strong coupling scales of
the theory, we can smoothly interpolate between the two dual
descriptions.  Here it is crucial that the gauge coupling going
through zero is a {\it smooth} process in the geometry: while the
K\"ahler volume of the 2-cycle vanishes as one goes through a flop,
the {\it physical} volume, given by \area, remains finite.
Moreover, we can read off from the geometry which
description is the more appropriate one at a given scale.

\subsec{The $A_2$ Example}

For illustration, we return to the example of the $A_2$ quiver studied
in section 3. To begin with, for a given set of charges $M_i$, we
take the couplings $g_i^{-2}$ of the theory to be {\it weak} at the scale
$\Delta$ set by the ``superpotential''. This is the characteristic
scale of the open-string ALE geometry. Then $S_i/\Delta^3$ is small in
the vacuum, and the weak coupling expansion is valid. From \run, we
can deduce the one-loop running of the couplings with energy scale
$\mu=t$
\eqn\betafn{
 \mu {d\over d \mu} g_{1}^{-2}(\mu) = (2 |M_1|+|M_3| - |M_2|),\qquad
 \mu {d\over d \mu} g_{2}^{-2}(\mu) = (2 |M_2|+|M_3| - |M_1|).
}
Suppose now, for example 
\eqn\cond{
2|M_1| + |M_3| \leq |M_2|,
}
so then at high enough energies, ${g_{1}^{-2}}(\mu)$ will become negative,
meaning that the size of $S^2_{i,t}$ has become negative. To keep the
size of all the $S^2$'s positive, at large enough $t$, the geometry
undergoes a flop of $S^2_1$ that sends
\eqn\flop{
\eqalign{
S^2_1 &\rightarrow \;\;\tilde{S}_1^2\,=\,- S^2_1\cr
S^2_2 &\rightarrow \;\; \tilde{S}_2^2\, =\, S^{2}_2 + S^2_1,
}
}
and correspondingly,  
\eqn\big{
{\tilde N}_1=N_2-N_1, \qquad {\tilde N}_2 = N_2,
}
while
\eqn\small{
{\tilde M}_{1} = -M_1, \qquad {\tilde M}_2=M_3, \qquad {\tilde
M}_3=M_2.  }
Recall the supersymmetric case first. The supersymmetric case with
$M_1=0$ was studied in detail in \vud . It corresponds to a vacuum of
a low energy $U(N_1)\times U(N_2)$, ${\cal N}=2$ theory where the
superpotential breaks the gauge group to $U(M_2)\times U(M_3)$.  The
formulas \betafn\ are in fact the same as in the supersymmetric case,
when all the $M_i$ are positive -- the beta functions simply depend
on the absolute values of the charges. If \cond\ is satisfied, the $U(N_1)$ factor is not asymptotically free, and the coupling grows strong at high energies. There, the theory is better described in terms of its Seiberg dual, the asymptotically free
$U({\tilde N}_1)\times U({\tilde N}_2)$
theory, broken to $U({\tilde M}_2)\times U({\tilde M}_3)$ by the
superpotential.\foot{The superpotential of the dual theory is not the
same as in the original. As explained in \vud , we can think of the
flop as permuting the $z'_i(t)$, in this case exchanging $z'_1(t)$
with $z'_2(t)$, which affects the superpotential as
$W_1(\Phi_1)\rightarrow-W_1(\Phi_1)$, and $W_2(\Phi_2)\rightarrow
W_1(\Phi_2)+W_2(\Phi_2)$.} The vacua at hand, which are visible
semi-classically in the $U(N_1)\times U(N_2)$ theory, are harder to
observe in the $U({\tilde N}_1)\times U({\tilde N}_2)$ theory, which is strongly coupled at the scale of the superpotential.  But, the duality predicts that they are there. In particular, we can smoothly vary
the strong coupling scale
$
{\Lambda}_{N_{1}}
$
of the original theory from $(i)$ 
${\Lambda}_{{N}_1}< \Delta< \mu$, where
the description at scale $\mu$ is better in terms of the
original 
$U({N}_1)\times U({N}_2)$ theory, to 
$(ii)$ $\Delta < \mu<{\Lambda}_{{N}_1}$, where
the description is better in terms of the dual
$U({\tilde N}_1)\times U({\tilde N}_2)$ theory.

For the dual description of a theory to exist, it is necessary, but not
sufficient (as emphasized in \Bena), that the brane charges at
infinity of the Calabi-Yau be the same in both descriptions. In
addition, the gauge couplings must run in a consistent way. In this
supersymmetric $A_2$ quiver, this is essentially true automatically,
but let's review it anyway with the non-supersymmetric case in
mind. On the one hand, \flopgc\ implies that the under the flop, the
couplings transform as
\eqn\match{{\eqalign{
g_{1}^{-2}(\mu)&\rightarrow {\tilde g}_{1}^{-2}(\mu)=
-{g}_{1}^{-2}(\mu)
\cr
g_{2}^{-2}(\mu)&\rightarrow 
{\tilde g}_{2}^{-2}(\mu)= {g}_{1}^{-2}(\mu)+ 
{g}_{2}^{-2}(\mu) .}}
}
On the other hand, from \run\ we know how the couplings ${\tilde
g}_{i}^{-2}$ corresponding to charges ${\tilde M}_i$ run with scale
$\mu$. The non-trivial fact is that the these two are consistent --
the flop simply exchanges ${\tilde M_2} =M_3$ and ${\tilde M}_3=M_2$,
and this is consistent with \match . 

Now consider the non-supersymmetric case. Let's still take $M_1=0$,
but now with $M_2>0>M_3$, such that \cond\ is satisfied. It is still
the case that if we go to high enough energies, i.e. large enough
$\mu$, the gauge coupling $g_1^{-2}$ will become negative, and the
corresponding $S^2_1$ will undergo a flop. We can change the basis of
2-cycles as in \flopgc\ and \flop\ so that the couplings are all
positive, and then the charges transform according to \small
. Moreover, just as in the supersymmetric theory, after the flop the
gauge couplings run exactly as they should given the new charges
${\tilde M}_i$, which are again obtained by exchanging node two and
three. Moreover, by varying the scale ${\Lambda}_{N_1}$ where
$g_1^{-2}$ becomes strong, we can smoothly go over from one
description to the other, just as in the supersymmetric case. For
example, in the $A_2$ case we have a non-supersymmetric duality
relating a
$
U(|N_1|)\times U(N_2)
$
theory, where the rank $N_1 = M_3$ is $negative$ and $N_2=M_2+M_3$ positive, which is a better description at low energies,
to a
$
U({\tilde N}_1) \times U({\tilde N}_2)
$ 
theory with positive ranks ${\tilde N}_1=N_2-N_1=M_2$ and ${\tilde
N}_2=N_2=M_2+M_3$, which is a better description at high energies.

More generally, one can see that this will be the case in any of the
ADE examples of the previous section. This is true regardless of
whether all $M_{I,p}$ are positive and supersymmetry is unbroken, or
they have different signs and supersymmetry is broken.  In the case
where supersymmetry is broken, we have no gauge theory predictions to
guide us, but it is still natural to {\it conjecture} the corresponding
non-supersymmetric dualities based on holography.  Whenever the charges are such that in going from low to high energies a root ends up being dualized
$$
S^2_{i,p}\rightarrow - S^2_{i,p},
$$
there should be a non-supersymmetric duality relating a
brane/anti-brane system which is a better description at low energies
to the one that is a better description at high energies, with charges
transforming as in \larggen\ and \smallgen. The theories are dual in
the sense that they flow to the same theory in the IR, and moreover,
there is no sharp phase transition in going from one description to
the other. This can be seen from the fact that by varying the strong
coupling scales of the theory, one can $smoothly$ interpolate between
one description and the other being preferred at a given energy scale
$\mu$. We don't expect these to correspond to gauge theory dualities
(in the sense of theories with a finite number of degrees of freedom
and a separation of scales), but we do expect them to be string theory
dualities.

\subsec{Dualizing an Occupied Root}

When an occupied node gets dualized, negative ranks $M<0$ will appear. This is true even in the supersymmetric case.  It is natural to wonder whether this is related to the appearance of non-supersymmetric vacua in a supersymmetric gauge theory. Conversely, starting with a non-supersymmetric vacuum at high energies, one may find that the good description at low energies involves all the charges being positive. We propose that when an occupied node gets dualized, there is essentially only one description which is ever really weakly coupled. In particular, ``negative rank'' gauge groups can appear formally but never at weak coupling. Moreover, while the supersymmetric gauge theories {\it can} have non-supersymmetric vacua, the phenomenon at hand is
unrelated to that. This is in tune with the interpretation given in
\vud .

Consider the $A_2$ theory in the supersymmetric case,
${\tilde M}_{1,2,3}>0$, with both gauge groups $U({\tilde N}_{1,2})$
being asymptotically free.  The $U({\tilde N}_1)\times U({\tilde N}_2)$ theory gives a good description at low energies, for
$$
\Lambda_{{\tilde N}_1} \ll \Delta
$$
where $\Delta$ is the characteristic scale of the ALE space, and $\Lambda_{{\tilde N}_1}$ is the strong coupling scale of the 
$U({\tilde N}_1)$ theory.  Now consider adiabatically increasing the strong coupling scale until 
$$
\Lambda_{{\tilde N}_1} \geq \Delta.
$$
Then the $U(N_1) \times U(N_2)$ description appears to be better at
low energies, with $N$'s related as in \big . Namely, from \match\ we
can read off the that the strong coupling scales match up as
$\Lambda_{N_1}=\Lambda_{{\tilde N}_1}$, so at least formally this
corresponds to a more weakly coupled, IR free $U(N_1)$ theory.
However, after dualizing node $1$, its charge becomes negative
$$ 
{\tilde M}_1=- M_1.  
$$ 
How is the negative rank $M_1<0$ consistent with the theory having a
supersymmetric vacuum? 

The dual theory clearly cannot be a weakly coupled theory.  A weakly
coupled theory of branes and anti-branes breaks supersymmetry, whereas
the solution at hand is supersymmetric. Instead, as we increase
${\Lambda}_{N_1}$ and follow what happens to the supersymmetric
solution, the scale ${\Lambda}_{M_1}$ associated with gaugino
condensation on node $1$ increases as well, $\Delta<
{\Lambda}_{M_1}\sim {\Lambda}_{N_1}$, and we find that at all energy
scales below $\Lambda_{M_1}$ we have a strongly coupled theory,
without a simple gauge theoretic description.
The holographic dual theory of course {\it does} have a weakly
coupled vacuum with charges $M_{1}<0,$ $M_{2,3}>0$, which 
breaks supersymmetry. However, the gauge couplings in this vacuum run 
at high energies in a different way than in the supersymmetric $U(N_1)\times U(N_2)$ gauge theory. As emphasized in \Bena , this means we {\it cannot} interpret this non-supersymmetric vacuum as a metastable state of the supersymmetric gauge theory.  

We could alternatively start with a weakly coupled, non-supersymmetric
$A_2$ theory with $M_1<0$, $M_{2,3}>0$. If \cond\ is not satisfied,
the theory is asymptotically free. Increasing the strong coupling
scale $\Lambda_{N_1}$ of this theory until $\Lambda_{N_1} \sim
\Delta$, the theory becomes strongly coupled, and one is tempted to
dualize it to a theory with ${\tilde M}_i>0$ at lower
energies. However, from the vacuum solutions in section 3, we can read
off that, just as in the supersymmetric case, this implies that the
scale $\Lambda_{M_1}$ of the gaugino condensate of node 1 becomes
larger than the scale $\Delta$, and no weakly coupled description
exists. What is new in the non-supersymmetric case is that, 
as we have seen in section 3, increasing the strong coupling scale $\Lambda_{M_1}$
to near $\Delta$ causes the system to lose stability.

Nevertheless, we can formally extend the conjectured Seiberg
dualities to all the supersymmetric and non-supersymmetric vacua even
when the node that gets dualized is {\it occupied}, except that the dual
description is, in one way or another, always strongly coupled.

\newsec{A Very Simple Case}

Let's now go back to the $A_2$ case studied in section 3 and
suppose that two of the masses are equal and opposite $m_1=-m_2=-m$,
so\foot{More precisely, relative to the notation of that section, we've performed a flop here that exchanges $z_1$ and $z_2$.}
\eqn\start{
z_1(t) =0, \qquad z_2(t) = -mt,\qquad z_3(t)=-m(t-a).
}
It is easy to see from \crit\ that there are now only two critical
points at $t=0$ and $t=a$, which get replaced by $S^2_1$ and
$S^2_3$. The third intersection point, which corresponds to the simple root
$S^2_2$, is absent here, and so is the minimal area 2-cycle
corresponding to it.  We study this as a special case since now the
prepotential ${\cal F}_0$ can be given in closed form, so the theory
can be solved exactly. This follows easily either by direct
computation from the geometry, or from the corresponding matrix
model (see appendix A). The large $N$ dual geometry corresponds to the two $S^2$'s being replaced by two $S^3$'s:
$$
x^2+y^2+z(z+mt)(z+m(t-a)) = s_1(z+ma)+s_3 (z+m(t-a)).
$$
The {\it exact} prepotential 
is given by
\eqn\presim{ 2 \pi i\, {\cal F}_0(S) = 
{1\over 2}S_1^2\, \Bigl(\log
({S_1\over m \Lambda_0^2})-{3\over 2}\Bigr) 
+{1\over 2}S_3^2\, \Bigl( \log
({S_3\over m \Lambda_0^2})-{3\over 2}\Bigr) 
+S_1S_3 \log({a\over
\Lambda_0}).
}
\bigskip
\centerline{\epsfxsize 5.0truein\epsfbox{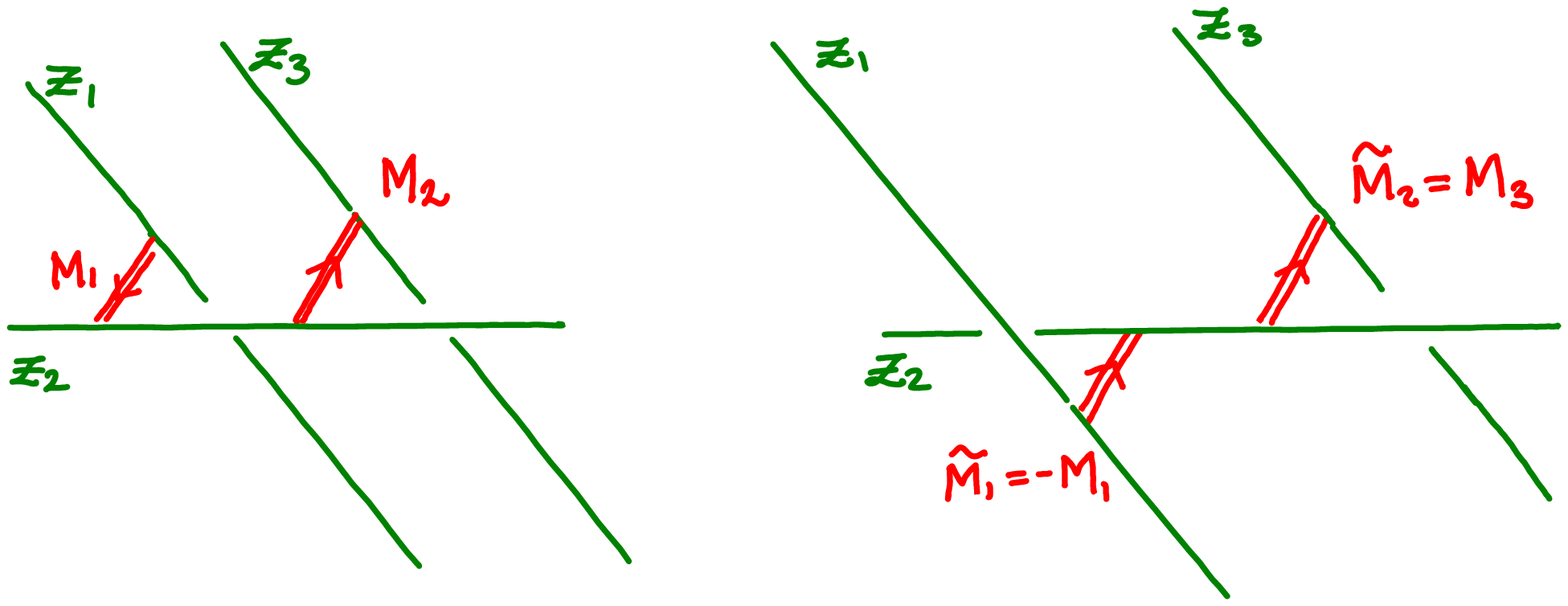}}
\noindent{\ninepoint
\baselineskip=2pt {\bf Fig. 3.} { There are only two minimal $S^2$'s in the $A_2$ geometry with $m_1=-m_2$. The
figure on the left corresponds to the first blowup discussed in the 
text, with two  minimal $S^2$'s of intersection number $+1$ in the ALE space wrapped by $M_1$ anti-D5 branes and $M_3$ D5 branes. The figure 
on the right is the flop of this.}}
\bigskip

We can now consider wrapping, say, $M_1$ anti-branes on $S^2_1$ and
$M_3$ branes on $S^2_3$.
We get an exact vacuum solution at
$$\eqalign{
{\overline{S}_1}^{|M_1|}& = 
\Bigl(\overline{\Lambda_0^2 m}\Bigr)^{|M_1|}
\Bigl({a\over \Lambda_0}\Bigr)^{-|M_3|}\exp(-2\pi i \alpha_1)
,\;\;\cr
{S_3}^{|M_3|} &= 
\Bigl({\Lambda_0^2 m}\Bigr)^{|M_3|}
\Bigl(\overline{a\over \Lambda_0}
\Bigr)^{-|M_1|}\exp(-2\pi i \alpha_3),}
$$
where the potential between the branes is given by
$$
V_* ={|M_1|\over g_{1}^2}+{|M_3|\over g_{3}^2}+
{1\over 2 \pi}|M_1||M_3|
\log\Bigl(|{a\over \Lambda_0}|\Bigr).
$$

Using an analysis identical to that in \Aganagic , it follows that 
the solution is {\it always} stable, at least in perturbation theory.
Borrowing results from \Aganagic ,
the masses of the four bosons 
corresponding to fluctuations of $S_{1,3}$ are given by
\eqn\masses{
\left( m_\pm (c) \right)^2 = {(a^2 +b^2 +2abcv)\pm \sqrt{(a^2 +b^2
+2abcv)^2 -4a^2b^2(1-v)^2} \over 2 (1-v)^2} 
}
and the masses of the corresponding fermions are
\eqn\fermasses{
|m_{\psi_1}|={a\over 1-v},\qquad |m_{\psi_2}|={b\over 1-v}
}
where $c$ takes values $c=\pm 1$,
and
\eqn\cont{a =\left|{M_1 \over 2\pi \Lambda_1^3 {\rm Im}
\tau_{11}}\right|, \qquad b = \left|{M_3\over 2\pi \Lambda_3^3
{\rm Im} \tau_{33}}\right|.  }
The parameter controlling the strength
of supersymmetry breaking $v$ is defined by
$$
v ={({\rm Im} \tau_{13})^2 \over {\rm Im} \tau_{11} {\rm Im}
\tau_{33}}.
$$
That $v$ controls
the supersymmetry breaking can be seen here from the fact that at
$v=0$, the masses of the four real bosons become degenerate in pairs, and
match up with the fermion masses \Aganagic .
The masses of bosons are strictly positive since the metric on the moduli space
${\rm Im} \tau$ is positive definite, which implies
$$\eqalign{
&1> v \geq 0\cr
&{\Lambda}_{N_{1,2}} \ll a
}$$
where ${\Lambda_{N_{1,2}}}$ is the scale at which the gauge coupling $g_{1,2}^{-2}$ becomes strong.\foot{From the solution, one can read off, e.g., $g_{1}^{-2}=- (2|M_1|+|M_3|)\log({\Lambda_{N_1}\over \Lambda_0}).$}

The fact that the system is stable perturbatively is at first sight surprising, since from the open string description one would expect that for sufficiently small $a$ an instability develops, ultimately related to the tachyon that appears when the brane separation is below the string scale. In particular, we expect the instability to occur when the coupling on the branes becomes strong enough that the Coulomb attraction overcomes the tension effects from the branes. However, it is easy to see that there is $no$ stable solution for small $a$.  As we decrease $a$, the solution reaches the boundary of the moduli space,
$$
\Lambda_0 \; \exp(-{1\over{ g_{1,3}^2| M_{3,1}|}})< a,
$$ 
where ${\rm Im}\tau$ is positive definite, {\it before} the instability can develop.\foot{Since ${\rm Im}(\tau)$ is a symmetric real matrix of rank two, a necessary condition for the eigenvalues to be positive is that the diagonal entries are positive. The equation we are writing corresponds to the positivity of the diagonal entries of ${\rm Im}(\tau)$ evaluated at the critical point. For weak gauge coupling, this is also the sufficient condition.}
Namely, if we view $\Lambda_0$ as a cutoff on how much energy one has available, then for a stable solution to exist at fixed coupling, the branes have to be separated by more than $\sim \Lambda_0$, and said minimum separation increases as one moves towards stronger coupling.  The couplings, however, do run with energy, becoming weaker at higher $\Lambda_0$, and because of that the lower bound on $a$ actually {\it decreases} with energy. Alternatively, as we will discuss in the next subsection, there is a lower bound on how small $|a|$ can get, set by the strong coupling scales $\Lambda_{N_{1,2}}$ of the brane theory. When this bound is violated, the dual gravity solution disappears.

The fact that the system is perturbatively stable should be related to the fact that in this case there is no compact $B$ cycle. Namely, in section 3 we have seen that when perturbative stability is lost, the system rolls down to a new minimum corresponding to shrinking a compact $B$-cycle without flux through it. In this case, such a compact $B$-cycle is absent, so the system has no vacuum it can roll away to, and correspondingly it remains perturbatively stable.

The theory has another vacuum with the same charges, which can have lower energy. This vacuum is not a purely closed string vacuum, but it involves branes. Consider, for example, the case with $M_1=-M_3=-M$.  In this case, the brane/anti-brane system should be exactly stable for large enough separation $a$. However, when $a$ becomes small enough, it should be energetically favorable to decay to a system with simply $M$ branes on $S^2_2$, which is allowed by charge conservation. This should be the case whenever
$$
A({S}^2_2)\leq 
A({S}^2_1)+A({S}^2_3)
$$
where the areas on the right hand side refer to those of the minimal ${S}^2$'s at the critical points of
${W}_1'(t) =
{z}_1(t)-{z}_2(t)$ and ${W}_3'(t) =
{z}_1(t)-{z}_3(t)$, 
$$
A({S}^2_1) = |r_1|, \qquad A({S}^2_3)= |r_1|+|r_2|.
$$
In the class of ${S}^2_2$, there is no holomorphic 2-cycle, as 
${W}_2'(t) = {z}_2(t)-{z}_3(t) = -ma$ 
never vanishes, so
\eqn\ar{
A({S}^2_2) = \sqrt{|r_2|^2 +|ma|^2}.
}
Clearly, when $a$ is sufficiently small, the configuration with $M$ branes on ${S}^2_2$ should correspond to the ground state of the system.  If instead $M_{1,3}$ are generic, we end up with a vacuum with intersecting branes, studied recently in \GiveonFK . Here one has additional massless matter coming from open strings at intersection of the branes, and correspondingly there is no gaugino condensation and no closed string dual. As a result, the methods based on holography we use here
have nothing to say about this vacuum.

\subsec{A Stable Non-Supersymmetric Vacuum}

Consider now the flop of the simple $A_2$ singularity of the previous sub-section, 
where $z_1$ and $z_2$ get exchanged,
$$
{\tilde z}_1(t) =-mt, \qquad {\tilde z}_2(t) =0,\qquad 
{\tilde z}_3(t)=-m(t-a),
$$
and where
$$
S_1^2 \rightarrow {\tilde S}^2_1= -S_1^2.
$$
We now wrap ${\tilde M}_1<0$ anti-branes on ${\tilde S}^2_1$ and ${\tilde M}_2>0$ branes on
${\tilde S}^2_2$. In this case, one would expect the system to have a stable, non-supersymmetric vacuum for $any$ separation between the branes. 
This is the case because the system has nowhere to which it can decay.
Suppose we wrap one anti-brane on ${\tilde S}^2_1$ and 
one brane on ${\tilde S}^2_2$. If a cycle $C$ exists such that
\eqn\cycle{
C = -{\tilde S}^2_1+{\tilde S}^2_2
}
then the brane/anti-brane system can decay to a brane on $C$. In the present case, such a $C$ does not exist. The reason for that is the following. On the one hand, all the curves in this geometry come from the ALE space fibration, 
and moreover all the $S^2$'s in the ALE space have self intersection number $-2$.
On the other hand, because the intersection number of ${\tilde S}^2_1$
and ${\tilde S}^2_2$ is $+1$, \cycle\ would imply that the self intersection of $C$ 
is $-6$. So, the requisite $C$ cannot exist.
The vacuum is, in fact, both perturbatively and non-perturbatively stable; we will see that the holographic dual theory has no perturbative instabilities 
for $any$ separation between the branes.

Because the $z$'s have been exchanged and the geometry is now different; 
we get a new prepotential ${\tilde {\cal F}}_0$ and effective superpotential
\eqn\news{\CW_{\rm eff} = 
\sum_{i=1,2} {\tilde \alpha}_i\; {\tilde S}_{i} + {\tilde
M}_{i}\; {\del_{{\tilde S}_{i}}} {\tilde {\cal F}}_0 ({\tilde S})
}
where 
$$ 2 \pi i\, {\cal F}_0({\tilde S}) = 
{1\over 2}{\tilde S}_1^2\, (\log
({{\tilde S}_1\over (-m) \Lambda_0^2})-{3\over 2})+ 
{1\over 2}{\tilde S}_2^2\, (\log
({{\tilde S}_2\over m \Lambda_0^2})-{3\over 2}) 
- {\tilde
S}_1{\tilde S}_2 \log({a\over\Lambda_0}). 
$$
Alternatively, we should be able to work with the old geometry and prepotential \presim,
but adjust the charges and the couplings consistently with the 
flop. The charges and the couplings of the two configurations are related by
\eqn\fr{
{\tilde M}_{1} = -M_1,\qquad {\tilde M}_2 = M_3,
}
where $M_{1,3}$ are now both positive, 
and 
\eqn\sn{
{\tilde g}^{-2}_1=- {g}^{-2}_1, \qquad
{\tilde g}^{-2}_2={g}^{-2}_3,
}
where ${g}^{-2}_1$ is now negative.  The effective superpotential is
\eqn\news{\CW_{\rm eff} = 
\sum_{i=1,3} {\alpha}_i\; {S}_{i} + {M}_{i}\; {\del_{{S}_{i}}} {\cal F}_0 ({S}),  
}
in terms of the old prepotential \presim .
Indeed, the two are related by ${\cal F}_{0}(S_1,S_3)={\tilde {\cal F}}_{0}({\tilde S_1,\tilde S_2})$
and a simple change of variables
$$
{\tilde S}_1 = -S_1,\qquad  {\tilde S}_2 = S_3,
$$
leaves the superpotential invariant.
The critical points of the potential associated to
\news\ with these charges
are
$$\eqalign{
&{\overline{{\tilde S}}_1}^{|{\tilde M}_1|} = \Bigl(\overline{\Lambda_0^2 m}
\Bigr)^{|{\tilde M}_1|}
\Bigl({{a\over \Lambda_0}}\Bigr)^{|{\tilde M}_2|}\, 
\exp(-2\pi i \alpha_1)\cr
&{{\tilde S}_2}^{|{\tilde M}_2|}\;\, = 
\Bigl({\Lambda_0^2 m}\Bigr)^{|{\tilde M}_2|}
\Bigl({\overline{a\over \Lambda_0}}\Bigr)^{|{\tilde M}_1|}\, 
\exp(-2\pi i \alpha_2)
}$$
with effective potential at the critical point
$$
V_* =
{|{\tilde M}_1|\over {\tilde g}_{1}^2}+{|{\tilde M}_2|\over {\tilde g}_{2}^2}-{1\over 2 \pi}|{\tilde M}_1||{\tilde M}_2|
\log\Bigl(|{a\over \Lambda_0}|\Bigr).
$$
The masses of the bosons in this vacuum are again given by
\masses\cont{} with the obvious substitution of variables.  Just as in
the previous subsection, the masses are positive in any of these
vacua. Moreover, because there are no two-loop corrections to the
prepotential, as we have seen in section 3, the vacuum is stable as
long as the metric remains positive definite. In the previous section,
we expected an instability for small enough $a$, and found that the
perturbatively stable non-supersymmetric solution escapes to the
boundary of the moduli space (defined as the region where ${\rm Im}
\tau$ is positive definite) when this becomes the case. In this case,
we do not expect any instability for any $a$, as there is nothing for
the vacuum to decay to. Indeed, we find that ${\rm Im} \tau$ is now
positive definite for any $a \neq 0$.

The vacuum is stable perturbatively and
non-perturbatively -- there simply are no lower energy states with the same charges available to which this can decay. So, this gives an example of an exactly stable, 
non-supersymmetric vacuum in string theory, albeit without four
dimensional gravity.\foot{This fact has been noted in \MukhiTE .} Moreover, since in this case there are no tachyons in the brane/anti-brane system, this should have a consistent limit where we decouple gravity and stringy modes, and are left with a pure, non-supersymmetric, confining gauge theory, with a large $N$ dual description. This is currently under investigation \toappear.

\centerline{\bf Acknowledgments}

We thank J. Heckman, K. Intriligator, S. Kachru, M. Mulligan, Y. Nomura, J. Seo, and especially C. Vafa for useful discussions.  The research of M.A., C.B., and B.F. is supported in part by the UC Berkeley Center for Theoretical Physics. The research of M.A. is also supported by a DOI OJI Award, the Alfred P. Sloan Fellowship, and the NSF grant PHY-0457317.

\appendix{A}{ Matrix Model Computation}

Using large $N$ duality in the B model topological string \dvone , the prepotential ${\cal F}_0$ of the Calabi-Yau manifolds studied in this paper can be computed using a matrix model describing branes on the geometry before the transition. The same matrix model \dvthree\ captures the dynamics of the  glueball fields $S$ in the ${\cal N}=1$ supersymmetric gauge theory in space-time, dual to the Calabi-Yau with fluxes in the physical superstring theory.  In this appendix, we use these matrix model/gauge theory techniques to compute the prepotential for Calabi-Yau manifolds which are $A_2$ fibrations with quadratic superpotentials, as studied in sections 3 and 6.  To our knowledge, this computation has not previously been carried out. 

The matrix model is a $U(N_1)\times U(N_2)$ quiver with Hermitian matrices $\Phi_1$ and $\Phi_2$ which transform in the adjoint of the respective gauge groups, and bifundamentals $Q$ and $\tilde{Q}$ which correspond to the bifundamental hypermultiplets coming from 12 and 21 strings.  The relevant matrix integral is then given by
$$
Z={1\over {\rm vol}\left(U(N_1)\times U(N_2)\right)}
\int d\Phi_1d\Phi_2dQd{\tilde Q}
\exp\left({1\over g_s}{}{\rm Tr}\CW(\Phi_1,\Phi_2,Q,{\tilde Q})\right)
$$
where $\CW$ is the superpotential of the corresponding ${\cal N}=1$ quiver gauge theory, given by
\eqn\mmss{
\CW=\Tr W_1(\Phi_1) + \Tr W_2(\Phi_2)+
\Tr({\tilde Q}\Phi_1{Q})-\Tr({Q}\Phi_2 {\tilde Q})
}
with
$$
\Tr W_1(\Phi_1)=-{m_1\over2}\Tr(\Phi_1 -a_1{\rm id}_{N_1})^2, \qquad
\Tr W_2(\Phi_2)=
-{m_2\over2}\Tr(\Phi_2-a_2 {\rm id}_{N_2})^2.
$$
The saddle points of the integral correspond to breaking the gauge group as 
\eqn\gauge{
U(N_1)\times U(N_2) \rightarrow U(M_1)\times U(M_2)\times U(M_3)
}
where 
$$
N_1=M_1+M_3,\qquad N_2=M_2+M_3,
$$ 
by taking as expectation values of the adjoints and bifundamentals to be
$$
\Phi_{1,*} = \pmatrix{
a_1\, {\rm id}_{M_1} & 0\cr
0 & a_3\, {\rm id}_{M_3}}, 
\qquad
\Phi_{2,*} = \pmatrix{
a_2\, {\rm id}_{M_2} & 0\cr
0 & a_3\, {\rm id}_{M_3}}
$$
where $a_3={(m_1 a_1 + m_2 a_2) /( m_1+m_2)}$, and
$$
(Q{\tilde Q})_* = \pmatrix{
0 & 0\cr
0 & -W_1'(a_3)\, {\rm id}_{M_3}},
\qquad
({\tilde Q}Q)_* = \pmatrix{
0 & 0\cr
0 & W_2'(a_3)\, {\rm id}_{M_3}},
$$
where $-W_1'(a_3) =m_1(a_1-a_3)=W_2'(a_3)$.

Now let's consider the Feynman graph expansion about this vacuum.  The end result is a very simple path integral.  However, to get there, we need to properly implement the gauge fixing \gauge , and this is somewhat laborious. It is best done in two steps. First, consider fixing the gauge that simply reduces $U(N_{1,2})$ to $U(M_{1,2})\times U(M_3)$.  This follows \dvg\ directly. Let
$$ 
\Phi_1 = 
\pmatrix{
\Phi^{1}_{11} & \Phi^{1}_{13} \cr
\Phi^{1}_{31} & \Phi^{1}_{33}}.
$$
To set the $M_1\times M_3$ block in $\Phi_1$ to zero
$$
F_1 = \Phi^{1}_{13} =0
$$
we insert the identity into the path integral in the form
$$
{\rm id} = \int 
d \Lambda \;\delta(F_1) \;{\rm Det}
({\delta F_1\over \delta \Lambda}),
$$
where the integral is over those gauge transformations not in $U(M_1)\times U(M_3)$. The determinant can be expressed in terms of two pairs of ghosts, $B_{13}, B_{31}$ and $C_{31}, C_{13}$, which are anti-commuting bosons, as
$${\eqalign{
{\rm Det}({\delta F_1\over \delta \Lambda}) 
=
\int dB_{13}dC_{31} d B_{31} d C_{13} 
\;& \exp\Bigl({1\over g_s}
\Tr( 
B_{13} \Phi^{1}_{33}C_{31} -  C_{31} \Phi^{1}_{11}B_{13})\Bigr)\cr
\;& \exp\Bigl({1\over g_s}\Tr(B_{31} \Phi^{1}_{11}C_{13}- C_{13} \Phi^{1}_{33}B_{31}  
)\Bigr).
}}
$$
By an identical argument, we can gauge fix the second gauge group factor 
$$
U(N_2)\rightarrow U(M_1)\times U(M_3)
$$
to set the $M_2\times M_3$ block of $\Phi_2$ to zero.  We do this by again inserting the identity into the path integral, but now with the determinant replaced by
$${\eqalign{
{\rm Det}({\delta F_2\over \delta \Lambda}) 
=
\int dB_{23}dC_{32} d B_{32} d C_{23} 
\;& \exp\Bigl({1\over g_s}
\Tr( 
B_{23} \Phi^{2}_{33}C_{32} -  C_{32} \Phi^{2}_{22}B_{23})\Bigr)\cr
\;& \exp\Bigl({1\over g_s}\Tr(B_{32} \Phi^{2}_{22}C_{23}- C_{23} \Phi^{2}_{33}B_{32}  
)\Bigr).
}}
$$

Finally, since the vacuum will break the two copies of $U(M_3)$ to a single copy, we need to gauge fix that as well. To do this, we'll fix a gauge 
$$ 
F_3 = Q_{33} - q \; {\rm id}=0 
$$ 
where $Q_{33}$ refers to the 33 block of $Q$, and $integrate$ over $q$. This is invariant under the diagonal $U(M_3)$ only. To implement this, insert the identity in the path integral, written as
$$
{\rm id} = \int d{\Lambda_{33}} \oint {dq\over q} \;
\delta(Q_{33} -q \, {\rm id}) \,q^{M_3^2}.
$$
The above is the identity since 
$$
{\rm Det}({\delta F_3 \over \delta \Lambda_{33}})=q^{M_{3}^2},
$$ 
and we have taken the $q$-integral to be around $q=0$.  Inserting this, we can integrate out $Q_{33}$, and ${\tilde Q}_{33}$.  The $Q_{33}$ integral sets it to equal $q$. The ${\tilde Q}_{33}$ integral is a delta function setting 
\eqn\set{
\Phi^1_{33} = \Phi^2_{33},
}
but there is a left over factor of $q^{-{M^2_3}}$ from the Jacobian of $\delta(q(\Phi^1_{33} - {\Phi^2}_{33}))$.  Integrating over $q$ gives simply $1$.  

The remaining fields include a pair of regular bosons $Q_{13}$,
${\tilde Q}_{31}$ in the bifundamental representation of $U(M_1)\times
U(M_3)$ and a pair of ghosts $C_{13}$, $B_{31}$, with {\it exactly} the
same interactions.  Consequently, we can integrate them out {\it exactly}
and their contribution is simply $1$. This also happens for $Q_{32}$,
${\tilde Q}_{23}$ and $B_{23}$, $C_{32}$, which also cancel out. We
are left with the spectrum presented in section 3 which very naturally describes branes with
open strings stretching between them.

\subsec{A Special Case}

In the special case when $m_2=-m_1=m$, the matrix integral is one-loop exact. To begin with, the effective superpotential is given by \mmss\ with
$$
\Tr W_1(\Phi_1)=-{m\over2}\Tr(\Phi_1)^2, \qquad
\Tr W_2(\Phi_2)=
{m\over2}\Tr(\Phi_2-a\; {\rm id}_{N_2\times N_2})^2.
$$
The theory now has only one vacuum, where $\Phi_1$ and $Q,$ 
${\tilde Q}$ vanish, and
$$
\Phi_2 = a \; {\rm id}_{N_2\times N_2}.
$$
Expanding about this vacuum, the superpotential can be re-written as
$${
\CW_{\rm eff} = -
{m\over 2} \Tr \,\Phi_{1}^2+{m\over 2} \Tr \,\Phi_{2}^2
-
a \Tr \,{Q}{\tilde Q}+
\Tr
( {\tilde Q}\Phi_{1}{Q}-{Q}\Phi_{2}{\tilde Q}).
}
$$
If we now redefine
$$
{\tilde \Phi}_1= \Phi_1 +{1\over m} Q {\tilde Q}, \qquad
{\tilde \Phi}_2= \Phi_2 +{1\over m} {\tilde Q}Q, \qquad
$$
the superpotential becomes quadratic in all variables, and the planar free energy is given by the exact expression:
$$
\CF_0= 
{S_1^2\over 2}\left(\log{S_1\over m\Lambda_0^2}-{3\over 2}\right)+
{S_2^2\over 2}\left(\log{S_2\over(-m)\Lambda_0^2}-{3\over 2}\right)
-
S_1S_2\log{a\over\Lambda_0}
$$
There are higher genus corrections to this result, but they all come from the volume of the $U(N)$ gauge groups, and receive no perturbative corrections.

\appendix{B}{ Geometrical calculation of the Prepotential}
One can derive the same prepotential by direct integration. We only sketch the computation here. The equation for the geometry \cs\ can be rewritten
\eqn\bgeom{\eqalign{
x^2 + y^2 + z (z - &m_1 (t - a_1))(z + m_2 (t - a_2)) = \cr
&-s_1 m_1 (z + m_2 (t - a_2)) -  s_2 m_2 (z - m_1 (t - a_1)) - s_3 m_3 z.
}}
Here $s_i$ are deformation parameters. This is a convenient rewriting of \cs\ because we will find that the periods of the compact cycles are given by $S_i = s_i + {\cal O}(S^2)$. As mentioned in the main text, the holomorphic three-form $\Omega$ of the Calabi-Yau descends to a one-form defined on the Riemann surface obtained by setting $x = y = 0$ in \bgeom. The equation for the Riemann surface is thus
\eqn\breim{
-1 = {m_1 s_1 \over  z (z - m_1 (t - a_1))} + {m_2 s_2 \over z (z + m_2 (t - a_2)) } 
+ {m_3 s_3 \over (z - m_1 (t - a_1))(z + m_2 (t - a_2)).}
}
The one-form can be taken to be $\omega = z dt - t dz$. The one-form is only defined up to a total derivative; a total derivative changes only the periods of the non-compact cycles, and our choice avoids quadratic divergences in the non-compact periods. These divergences would not contribute to physical quantities in any case. The equation for the Riemann surface is a cubic equation for $z(t)$, so the Riemann surface has three sheets, which are glued together along branch cuts. The compact periods are given by integrals around the cuts, while the non-compact periods are given by integrals from  the cuts out to a cutoff, which we take to be $t = \Lambda_0$.

It is convenient to make the change of variables
\eqn\uvvar{
u = {-t + a_1 + z/m_1 \over a_{21}} \ \ \ \ \ v = -z { m_3 \over a_{21} m_1 m_2}
}
where $a_{21} = a_2 - a_1$. In the new variables, the equation for the Riemann surface takes the simple form
\eqn\uvgeom{
1 =  {s_1 \over \Delta^3}{1 \over u v} - {s_2 \over \Delta^3} { 1 \over v (u + v + 1)}  + {s_3 \over \Delta^3}{ 1 \over u (u+v+1)}
}
with $\Delta^3 =  (a_2 - a_1)^2m_1 m_2/ m_3$ as in the main text. The change of variables is symplectic up to an overall factor, so in the new variables the one-form becomes
\eqn\uvoneform{
\omega = \Delta^3 (u dv - v du).
}
The change of variables makes it clear that we can think of the problem as having one dimensionful scale $\Delta$, and three dimensionless quantities, $s_i /\Delta^3$, which we will take to be small. There are many other dimensionless quantities in the problem, such as $m_i/m_j$, but they do not appear in the rescaled equations so they will not appear in the periods, with one small caveat. While the equation for the Riemann surface and the one-form only depend on $\Delta$ and $s_i /\Delta^3$, the cutoff is defined in terms of the original variables, $t = \Lambda_0$, so the cutoff dependent contributions to the periods can depend on the other parameters.

We sketch how to compute one compact period and one non-compact period. Though it is not manifest in our equations, the problem has a complete permutation symmetry among 
$(s_1, s_2, - s_3)$, so this is actually sufficient. One compact cycle (call it $S_1$) is related to the region in the geometry where $u$ and $v$ are small, so that to a first approximation
\eqn\bapprox{
1 \approx  {s_1 \over \Delta^3}{1 \over u v}.
}
We expand \uvgeom\ for small $u,v$ to get
\eqn\gapsone{
u v =  {s_1 \over \Delta^3} -  {s_2 \over \Delta^3} u (1 - u - v) +  {s_3 \over \Delta^3}v (1 - u - v) + ...
}
This will be sufficient for the order to which we are working, and the equation is quadratic. 
We could solve for $u(v)$ or $v(u)$ in this regime; we would find a branch cut and integrate the one-form around it. Equivalently, we can do a two dimensional integral
\eqn\bsone{
S_1 = \Delta^3 \int du \wedge dv
}
over the region bounded by the Riemann surface (this is Stokes' Theorem). One can derive a general formula for the integral over a region bounded by a quadratic equation by changing coordinates so that it is the integral over the interior of a circle. In this case, the result is
\eqn\gppsone{
S_1 = s_1 + {1 \over \Delta^3} \left(s_1 s_2 - s_1 s_3 - s_2 s_3 \right) +  {\cal O}\left(s^3 \over \Delta^6 \right).
} 
The permutation symmetry of the problem then determines the other compact periods.

Now we compute the integral over the cycle dual to $S_1$. The contour should satisfy $ uv \approx s_1 /\Delta^3$ and go to infinity. Also, the contour must intersect the compact 1-cycle in a point. A contour which satisfies these criteria is to take $u,v$ to be real and positive (this choice works as long as the $s_i$ are real and positive, but the result will be general). We will need two different perturbative expansions to do this integral: one for small $u$ and the other for small $v$. Since we have $u v \approx s_1/\Delta^3$, we will need a ``small $u$" expansion which is valid up to $u \sim \sqrt{s_1/ \Delta^3}$, and similarly for the small $v$ expansion. 

To expand for small $v$, we first multiply \uvgeom\ through by $v$ to get
$$    
v = {s_1 \over \Delta^3}{1 \over u } - {s_2 \over \Delta^3} { 1 \over  1 + u + v}  + {s_3 \over \Delta^3}{ v \over u (1 + u+v)}
$$
We now solve perturbatively for $v(u)$, using the fact that throughout the regime of interest $v << 1 + u$. The largest that $v/(1+u)$ gets in this regime is 
$$
{v \over 1+u} < \sqrt{s_1 \over \Delta^3}.
$$
To zeroth order in $v/(1+u)$,
\eqn\vzero{
v^{(0)} = {s_1 \over \Delta^3}{1 \over u }- {s_2 \over \Delta^3}{ 1 \over  1 + u}.
}
To first order,
$$
v^{(1)} =  {s_1 \over \Delta^3}{1 \over u } - {s_2 \over \Delta^3} { 1 \over  1 + u + v^{(0)}}
+ {s_3 \over \Delta^3}{ v^{(0)} \over u (1 + u)},
$$
which upon expanding becomes
\eqn\vone{
v^{(1)} = {s_1 \over \Delta^3}{1 \over u }  - {s_2 \over \Delta^3} \left({ 1 \over  1 + u}  -  {v^{(0)}  \over (1 + u)^2}\right)+ {s_3 \over \Delta^3}{ v^{(0)} \over u (1 + u)}.
}
We need keep one more order in the perturbative expansion in order to get the prepotential to the desired order:
$$
v^{(2)} =  {s_1 \over \Delta^3}{1 \over u } - {s_2 \over \Delta^3} { 1 \over  1 + u + v^{(1)}}  + {s_3 \over \Delta^3}{ v^{(1)} \over u (1 + u+v^{(0)})}
$$
which upon expanding becomes
\eqn\vtwo{
v^{(2)} =  {s_1 \over \Delta^3}{1 \over u } 
- {s_2 \over \Delta^3}\left[ { 1 \over  1 + u } - {v^{(1)} \over (1+u)^2} + {(v^{(0)})^2 \over (1+u)^3} \right]
+ {s_3 \over \Delta^3}\left[{ v^{(1)} \over u (1 + u)} - {(v^{(0)})^2 \over (1+u)^2}\right]
}
Note that using \vzero\vone, this is an explicit equation for $v(u)$.
We could similarly expand to find $u(v)$ in the regime of small $u$, but actually we can save ourselves the computation by noting that the equation for the Riemann surface is invariant under $u \leftrightarrow v, s_2 \leftrightarrow -s_3$. We are now in a position to perform the integral of the one-form $\omega$ over the contour. We use the approximation \vtwo\ for the part of the integral where $v$ is small, and the corresponding formula for $u(v)$ for the part of the integral where $u$ is small. We can choose to go over from one approximation to another at a point $u_{\min} = v_{\min}$. Such a point will be approximately
$u_{\min} = \sqrt{s_1/\Delta^3}$, but we need a more precise formula. By setting $u=v$ in the equation for the Riemann surface, and perturbing around $u_{\min} = \sqrt{s_1/\Delta^3}$, we find
$$
u_{\min}^2 = {s_1 \over \Delta^3} -{s_2 - s_3 \over \Delta^3} \sqrt{s_1 \over \Delta^3}  + {(s_2 - s_3)^2 \over 2 \Delta^6} + {2 s_1 (s_2 - s_3) \over \Delta^6} + \ldots
$$
We spare the reader the details of the integration. The result is cutoff dependent, and we assume that the cutoff is sufficiently large so that we can drop contributions which depend inversely on the cutoff. After doing the integral, we rewrite the result in terms of the compact periods $S_i$ using \gppsone.
The result is:
$$\eqalign{
\partial_{S_1} {\cal F}_0 = (S_1 &- S_2)\log u_{\max} + ( S_1 +  S_3)\log v_{\max} - ( S_1 \log {S_1 \over \Delta^3} - S_1)\cr
&- {1 \over \Delta^3} \left({1\over 2}S_2^2 
+ {1 \over 2} S_3^2 + S_1 S_2 - S_1 S_3 - 3 S_2 S_3 \right)  + {\cal O}\left(S^3 \over \Delta^6\right).
}$$
Here $u_{\max}$ and $v_{\max}$ are cutoffs at large $u, v$. Since our cutoff is $t=\Lambda_0$, we can solve for $u_{\max}, v_{\max}$. When $u$ is large, $v$ is small, since $uv \approx S_1/\Delta^3$. Looking back at the change of variables \uvvar, we find
$$
u_{\max} = {\Lambda_0 \over a_{21}} \ \ \ \ \ \ \ \ \ \ \ \
v_{\max} = {\Lambda_0 m_3 \over a_{21} m_2} = {\Lambda_0 \over a_{31}}
$$
Again, the other non-compact periods are determined by symmetry. It is now a simple matter to find ${\cal F}_0$:
$$\eqalign{
2 \pi i {\cal F}_0 &= \half S_1^2 \log {\Lambda_0^2 \over a_{21} a_{31}}
+ \half S_2^2 \log {\Lambda_0^2 \over a_{21} a_{23}}
+  \half S_3^2 \log {\Lambda_0^2 \over a_{31} a_{23}}\cr
&- S_1 S_2 \log{\Lambda_0 \over a_{21}} 
+ S_1 S_3 \log{\Lambda_0 \over a_{31}} 
+ S_2 S_3 \log{\Lambda_0 \over a_{23}}\cr
&- \half S_1^2 \left(\log {S_1 \over \Delta^3} - {3 \over 2} \right)
- \half S_2^2 \left(\log {S_2 \over \Delta^3} - {3 \over 2} \right)
- \half S_3^2 \left(\log {S_3 \over \Delta^3} - {3 \over 2} \right)\cr
&- {1 \over 2\Delta^3} \left(S_1 S_2^2 + S_1^2 S_2 + S_1 S_3^2 - S_1^2 S_3 + S_2 S_3^2 - S_2^2 S_3 - 6 S_1 S_2 S_3 \right) + {\cal O} \left({S^4 \over \Delta^6}\right).
}$$ 
This result agrees with the matrix model computation of appendix A. Recall that we dropped terms which depend inversely on the cutoff. More precisely, we dropped contributions to the non-compact period of the form $S_i |a_{12}|/\Lambda_0$. This is necessary in order to match the result of the matrix model computation. In particular, in order to justify keeping the corrections we do keep, we require
\eqn\fddd{
{S_i \over \Delta^3} >> {|a_{12}| \over \Lambda_0}.
}

\appendix{C}{ The Hessian at Two Loops}

The equations required to analyze stability simplify if we introduce the notation
$$
u^a \equiv i G^{ab}\alpha_b.
$$ 
Since we are taking the $\alpha_i$ to be pure imaginary, $u^a$ will be real and positive. Furthermore, since we are taking $\tau_{ab}$ to be pure imaginary, we can replace it with the metric, $\tau_{ab} = i G_{ab}$. Then the equation of motion \twoeom\ takes the simple form 
\eqn\twosimsim{
{1 \over 2} i {\cal F}_{kab} (u^a u^b - M^a
M^b) = 0.
}
At one-loop, the third derivative of the prepotential is nonzero only if all of the derivatives are with respect to the same variable, so at one-loop the solutions are
$
u^a = \pm M^a.
$
As discussed earlier, the physically relevant solutions are
\eqn\twophysone{
u^a = |M^a|.
}
This is just a rewriting of the one-loop solutions \so\ in terms of the new notation. 

At two loops, we can find the solution by perturbing around the one-loop result. Let $u^a = |M^a| + \delta^a $. We find that
\eqn\twodeltsolsol{
\delta^k = {1 \over 2 |M^k|{\cal F}_{kkk}} {\cal F}_{kab}(- |M^a| |M^b| + M^a M^b).
}
Having solved the equations of motion at two loops, we proceed to the Hessian, providing less detail. Assuming the same reality conditions, the matrices of second derivatives are given by
\eqn\twov{
\partial_k \partial_l V = \partial_{\bar k} \partial_{ \bar l} V = 
{1 \over 2} 
\left( i {\cal F}_{abkl} + i{\cal F}_{cak}i{\cal F}_{dbl}G^{cd}
\right)
(u^a u^b - M^a M^b),
}
\eqn\twovmixed{
\partial_k \partial_{\bar l} V = \partial_{\bar k} \partial_l V = 
{1 \over 2} i{\cal F}_{cak}i{\cal
F}_{dbl} G^{cd}(u^a u^b + M^a M^b).
}
The relations between the different mixed partial derivatives arise because we are perturbing about a real solution. 

At two loops, taking four derivatives of the prepotential gives zero unless all of the derivatives are with respect to the same variable, so the first term in \twov\ can be simplified as
\eqn\twosimp{
 i{\cal F}_{abkl}(u^a u^b - M^a M^b)= \delta_{kl} i {\cal
 F}_{kkkk}(u^k u^k - M^k M^k)=2 \delta_{kl} i {\cal
 F}_{kkkk} |M^k| \delta^k.
}
Though it is not obvious at this stage, the other terms on the right hand side can be approximated by their one-loop value in the regime of interest. This is very useful because, as mentioned previously, at one-loop the third derivatives of the prepotential vanish unless all indices are the same. With these simplifications, the nonzero second derivatives become
\eqn\tworealrealfact{\eqalign{
(\partial_a + \partial_{\bar a})(\partial_b +
\partial_{\bar b}) V &= \cr 
\sum_c 2 i{\cal F}_{aaa}&i{\cal F}_{ccc}G^{ac} |M^a| |M^c| 
\left(\delta_{cb} + 
 G_{cb} {i{\cal F}_{bbbb} \delta^b \over  
i{\cal F}_{bbb}i{\cal F}_{ccc}  |M^c|} \right),
}} 
\eqn\twoimimfact{\eqalign{
\,\;\;\;\;\;(\partial_a - \partial_{\bar a})(-\partial_b +
\partial_{\bar b}) V &= \cr  
\sum_c 2 i{\cal F}_{aaa}&i{\cal F}_{ccc}G^{ac} |M^a| |M^c| 
\left(\delta_{cb} - 
 G_{cb} {i{\cal F}_{bbbb} \delta^b \over  
i{\cal F}_{bbb}i{\cal F}_{ccc}  |M^c|} \right).
}} 
In these equations, no indices are implicitly summed over. 

In order to analyze the loss of perturbative stability, we compute the determinant of the Hessian. Since the eigenvalues remain real, in order to go from a stable solution to an unstable one, an eigenvalue should pass through zero. We therefore analyze where the determinant is equal to zero. Up to possible constant factors, the determinant is given by
\eqn\twodetmm{
\left( {\rm Det} \  G^{ab}\right)^2 \left(\prod_c {|M^c| \over i {\cal
F}_{ccc}}\right)^4 {\rm Det}\left(\delta_{cb} + 
 G_{cb} {{\cal F}_{bbbb} \delta^b \over  
i{\cal F}_{bbb}{\cal F}_{ccc} |M^c|} \right)
{\rm Det}\left(\delta_{cb} - 
 G_{cb} {{\cal F}_{bbbb} \delta^b \over  
i{\cal F}_{bbb}{\cal F}_{ccc}  |M^c|} \right)
}
and so in order to vanish, one of the last two determinants must go to zero.
\listrefs
\end